\documentclass[prx,aps,twoside,twocolumn,floatfix,showpacs,superscriptaddress]{revtex4}

\usepackage{latexsym}
\usepackage{graphicx}
\usepackage{amsmath,amssymb,hyperref}
\usepackage[squaren]{SIunits}
\usepackage{bm}
\usepackage{tabularx}
\usepackage{url}
\usepackage{color}
\definecolor{pink}{RGB}{255,0,255}  

\newcommand {\ket}[1] {|#1 \rangle}
\newcommand {\densop}[2] {|#1 \rangle \langle #2 |}

\begin{document}
	\title{Space QUEST mission proposal: Experimentally testing decoherence due to gravity}

	\author{Siddarth Koduru Joshi}
	\affiliation{Institute for Quantum Optics and Quantum Information, 
		Austrian Academy of Sciences, Boltzmanngasse 3, A-1090 Vienna, Austria}    
	
	\author{Jacques Pienaar}
	\affiliation{Institute for Quantum Optics and Quantum Information, 
		Austrian Academy of Sciences, Boltzmanngasse 3, A-1090 Vienna, Austria}
	
	\author{Timothy C. Ralph}
	\affiliation{Centre for Quantum Computation \& Communication Technology, Department of Physics, The University of Queensland, St Lucia QLD 4072}
	
	\author{Luigi Cacciapuoti}
	\affiliation{European Space Agency, Keplerlaan 1 - P.O. Box 299, 2200 AG Noordwijk ZH, The Netherlands}
	
	\author{{Will McCutcheon}}
	\affiliation{Department of Electrical and Electronic Engineering,
		University of Bristol, Merchant Venturers Building, Woodland Road, Bristol, BS8 1UB, UK}
	
	\author{{John Rarity}}
	\affiliation{Department of Electrical and Electronic Engineering,
		University of Bristol, Merchant Venturers Building, Woodland Road, Bristol, BS8 1UB, UK}
	
	\author{{Dirk Giggenbach }}
	\affiliation{German Aerospace Center, Institute of Communications and Navigation, Muenchner Strasse 20, 82234 Wessling, Germany}
	
	\author{Jin Gyu Lim}
	\affiliation{Institute for Quantum Computing and Department of Electrical and Computer Engineering, University of Waterloo, Waterloo, ON, N2L~3G1 Canada}
	
	\author{Vadim Makarov}
	\affiliation{Institute for Quantum Computing and Department of Physics and Astronomy, University of Waterloo, Waterloo, ON, N2L~3G1 Canada}

	\author{Ivette  Fuentes}\affiliation{Institute for Quantum Optics and Quantum Information, 
		Austrian Academy of Sciences, Boltzmanngasse 3, A-1090 Vienna, Austria}

	\author{Thomas Scheidl}
	\affiliation{Institute for Quantum Optics and Quantum Information, 
		Austrian Academy of Sciences, Boltzmanngasse 3, A-1090 Vienna, Austria}
	
	\author{Erik Beckert}
	\affiliation{Precision Engineering Department, Fraunhofer IOF, Albert-Einstein-Stra\ss{}e 7, 07745 Jena, Germany}
	\author{ Mohamed  Bourennane}
	\affiliation{Physics department, Stockholm University, Albanova universitetscentrum, Universitetsv\"{a}gen 10, 114 18 Stockholm, Sweden}
	
	\author{David Edward Bruschi}
	\affiliation{York Centre for Quantum Technologies, Department of Physics, University of York, YO10 5DD Heslington, UK}
	
	\author{Adan  Cabello}
	\affiliation{University of Seville, E-41012 Sevilla,
		Spain}
	
	\author{ Jose  Capmany}
	\affiliation{iTEAM Research Institute, Universitat Politècnica de Valencia, Valencia 46022, Spain.}
	
	
	\author{Alberto  Carrasco-Casado}
	\affiliation{Space Communications Laboratory,
		National Institute of Information and Communications Technology (NICT),
		4-2-1, Nukui-Kitamachi,
		184-8795 Koganei,
		Tokyo, Japan}
	
	\author{Eleni  Diamanti}
	\affiliation{ CNRS, Universite Pierre et Marie Curie, 75005 Paris, France}
	
	\author{ Miloslav  Du\u{s}ek}
	\affiliation{Department of Optics, Faculty of Science, Palacky University, 17. Iistopadu 12, 772 00, Olomouc, Czech Republic}
	
	\author{Dominique  Elser}
	\affiliation{Max Planck Institute for the Science of Light, G\"{u}nther-Scharowsky Str. 1 Bldg. 24, D-91058 Erlangen, Germany}
	
	\author{Angelo  Gulinatti}
	\affiliation{Politecnico di Milano, Dipartimento di Elettronica, Informazione e Bioingegneria (DEIB) Piazza Leonardo da Vinci 32, 20133 Milano, Italy}
	
	\author{ Robert H. Hadfield}
	\affiliation{School of Engineering, University of Glasgow, Glasgow G12 8LT, United Kingdom}
	
	\author{ Thomas  Jennewein} 
	\affiliation{Institute for Quantum Computing and Department of Physics and Astronomy, University of Waterloo, Waterloo, ON, N2L~3G1 Canada}
	
	\author{Rainer  Kaltenbaek }
	\affiliation{Vienna Center of Quantum Science and Technology, Faculty of Physics, University of Vienna, Boltzmanngasse 5, A-1090 Vienna, Austria}
	
	\author{Michael A. Krainak}
	\affiliation{NASA Goddard Space Flight Center
		Greenbelt, MD 20771}
	
	\author{ Hoi-Kwong  Lo}
	\affiliation{Center for Quantum Information and Quantum Control,
		Department of Physics and Department of Electrical and Computer Engineering,
		University of Toronto, M5S 3G4 Toronto, Canada}
	
	\author{Christoph  Marquardt}
	\affiliation{Max Planck Institute for the Science of Light, G\"{u}nther-Scharowsky Str. 1 Bldg. 24, D-91058 Erlangen, Germany}
	
	\author{Gerard  Milburn}
	\affiliation{Department of Physics, University of Queensland, St Lucia, QLD 4072, Australia}
	
	\author{ Momtchil  Peev}
	\affiliation{Quantum Technologies, Smart Systems Division, Austrian Research Centers GmbH ARC, Donau-City, Strasse 1, 1220 Vienna, Austria}
	
	\author{Andreas  Poppe}
	\affiliation{Safety and Security Department, AIT Austrian Institute of Technology GmbH, 2444 Seibersdorf, Austria}
	
	\author{Valerio  Pruneri}
	\affiliation{ICFO\textendash{}Institut de Ci\`{e}ncies Fot\`{o}niques, The Barcelona Institute of Science and Technology, 08860 Barcelona, Spain, and Instituci\'{o} Catalana de Recerca i Estudis Avan\c{c}ats, 08010 Barcelona, Spain}
	
	\author{Renato  Renner}
	\affiliation{Institute for Theoretical Physics, ETH Zurich, 8093 Switzerland}
	
	\author{Christophe  Salomon}
	\affiliation{Laboratoire Kastler Brossel, ENS-PSL Research University, CNRS, UPMC, Coll`ege
		de France, 24, rue Lhomond, 75005 Paris}
	
	\author{ Johannes  Skaar}
	\affiliation{Department of Electronics and Telecommunications,
		Norwegian University of Science and Technology, NO-7491 Trondheim, Norway}
	
	\author{ Nikolaos  Solomos}
	\affiliation{Physical Sciences Sector, Applied Physics \& Naval Electrooptics  Laboratories, Hellenic Naval Academy, Hatzikyriakeion, Piraeus, 18503, Greece}
	
	\author{Mario Stip\v{c}evi\'c}
	\affiliation{Ruder Bo\v{s}kovi\'{c} Institute, Center of Excellence for Advanced Materials and Sensing Devices and Division of Experimental Physics, 10000 Zagreb, Croatia}
	
	\author{ Juan P. Torres}
	\affiliation{ICFO- Institut de Ciencies Fotoniques, Barcelona Institute of Science and Technology, and Universitat Politecnica de Catalunya 08860 Barcelona, Spain}
	
	\author{ Morio  Toyoshima}
	\affiliation{Space Communications Laboratory,
		National Institute of Information and Communications Technology (NICT),
		4-2-1, Nukui-Kitamachi,
		184-8795 Koganei,
		Tokyo, Japan}
	
	\author{ Paolo  Villoresi}
	\affiliation{Dipartimento di Ingegneria dell’Informazione, Universit\`a degli Studi di Padova, Padova, Italy}
	
	\author{Ian  Walmsley}
	\affiliation{Clarendon Laboratory, University of Oxford, Parks Road, Oxford OX1 3PU, United Kingdom}
	
	\author{Gregor  Weihs}
	\affiliation{Institut f\"{u}r Experimentalphysik, Universit\"{a}t Innsbruck, Technikerstr. 25, 6020 Innsbruck, Austria}
	
	\author{ Harald  Weinfurter}
	\affiliation{Max-Planck-Institut f\"{u}r Quantenoptik, Hans-Kopfermann-Straße 1, 85748 Garching, Germany}
	
	\author{ Anton  Zeilinger}
	
	\affiliation{Institute for Quantum Optics and Quantum Information,     Austrian Academy of Sciences, Boltzmanngasse 3, A-1090 Vienna, Austria}
	
	\author{ Marek  \.{Z}ukowski}
	
	\affiliation{Institute of Theoretical Physics and Astrophysics, University of Gda\'{n}sk,  80-308 Gda\'{n}sk, Poland}
	
	\author{Rupert Ursin}
	\affiliation{Institute for Quantum Optics and Quantum Information,     Austrian Academy of Sciences, Boltzmanngasse 3, A-1090 Vienna, Austria}
	
	\collaboration{Space QUEST topical team}
	
	\date{\today}

	\begin{abstract}

		Models of quantum systems on curved space-times lack sufficient experimental verification. Some speculative theories suggest that quantum properties, such as entanglement, may exhibit entirely different behavior to purely classical systems. By measuring this effect or lack thereof, we can test the hypotheses behind several such models. For instance, as predicted by Ralph and coworkers [T C Ralph, G J Milburn, and T Downes, Phys. Rev. A, 79(2):22121, 2009; T  C  Ralph  and  J  Pienaar, New Journal of Physics, 16(8):85008, 2014], a bipartite entangled system could decohere if each particle traversed through a different gravitational field gradient.
		We propose to study this effect in a ground to space uplink scenario. We extend the above theoretical predictions of  Ralph and coworkers and discuss the scientific consequences of detecting/failing to detect the predicted gravitational decoherence. We present a detailed mission design of the European Space Agency's (ESA) Space QUEST (Space - Quantum Entanglement Space Test) mission, and study the feasibility of the mission scheme. 
	\end{abstract}
	
	
	\maketitle

\section{Introduction} \label{sec:intro}

Consider a quantum mechanical system consisting of two entangled photons. One photon of each pair is detected on the ground while the other is sent to the International Space Station (ISS).
Different theoretical models have been proposed to analyze this scenario with widely varying results. For example, one possible approach would be to place the system on a curved background metric and quantise the field over the set of modes formed by the geodesics on the metric. Thus taking into account  the differences in expected arrival times (due to path lengths, turbulence, time-dilation, clock drifts, etc.). Standard Quantum Mechanics (QM) predicts no additional decoherence due to the difference in gravitational curvature between the two photon paths. Such an approach, however, breaks down in the presence of exotic gravitational fields with non-hyperbolic metrics. So one requires new models to deal with these types of space-time curvatures. Another analysis using quantum field theory in curved space-time shows that a single-photon wave-packet undergoes not only a Doppler frequency shift, but also the broadening of the mode profile. This broadening occurs because the propagation through curved space-time induces an effective change in refractive index that shifts excitations to other frequency modes~\cite{bruschi2014spacetime,bruschi2014quantum,kohlrus2015quantum}. There have also been different predictions for a gravitationally induced decoherence effect. Reference~\cite{bruschi2014spacetime} shows that a decoherence effect is produced by the shifting and the flattening of the wavepacket's frequency distribution. In this derivation, no particle creation was assumed to happen. Thus, it is  possible, in principle, to recover the information that has been lost by adjusting the detector to correct the gravitationally induced effects. In contrast, the model proposed in Refs.~\cite{RAL14,RMD09} predicts an irrecoverable gravitational decoherence effect due to a speculative nonlinear back-action of the metric on the quantum fields that leads to particle loss into a causally disconnected region of space-time. Furthermore, Refs.~\cite{RMD09,RAL14,ANA13,WAN,BRE,KAY,DAS08,PIK12} show that this type of decoherence effect is seen only by entangled systems (i.e.,\ classical correlations are not affected).

Uniquely, the predictions of Refs.~\cite{RMD09,RAL14}, referred to as the event operator formalism, can be experimentally verified with current technology, providing a rare opportunity to test {models of} fundamental QM and general relativity simultaneously.  We present (in Section~\ref{sec:theory}) an extension of the theoretical framework behind the mission, to address practical concerns {including} losses and the need for space-like separation of detection events. We also show that certain types of entangled quantum systems/states show a more pronounced decoherence effect and study the feasibility of a {simple, low-cost} space-based mission to search for decoherence in quantum systems due to gravity (Section~\ref{sec:feas}). In Section~\ref{sec:expt} we present a minimalistic experimental design which utilizes several resources already on board the ISS. Further, the same apparatus can be used for many other far-reaching quantum experiments including long distance Bell tests, {a variety of } Earth-to-space quantum communication protocols, ~\cite{Scheidl2013a}. The same flight hardware can also be used for classical communication research; for example, the precise time tags can be used for very high order ($\approx$2048) Pulse Position Modulation (PPM)~\cite{ppm}, while the polarization channels can be used for Multiple Input Multiple Output Transmit Diversity~\cite{polmux}.

The Experiment Scientific Requirement (ESR) document was submitted to the European Space Agency (ESA) at the conclusion of the Phase-A study. The ESR defines the scientific objectives and derives from them the requirements driving the mission design, while this manuscript furthers the underlying science as well as details the feasibility study and one recommended implementation on the ISS.

\section{Theory} \label{sec:theory}

\begin{figure}[!htbp]
	\includegraphics[width=0.85\columnwidth]{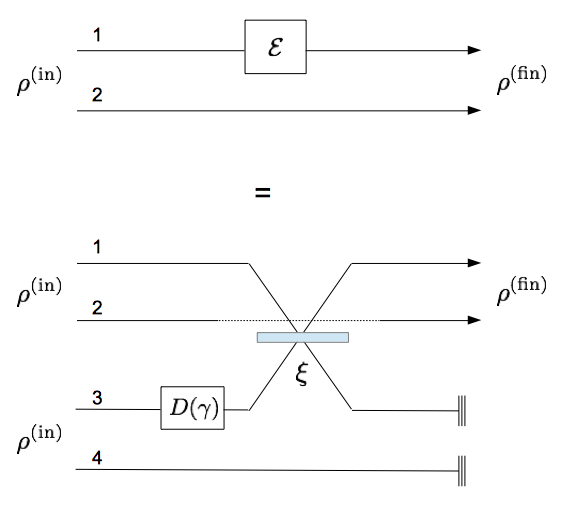}
	\caption{The event operator formalism can be understood as a nonlinear map $\mathcal{E}$ acting on the mode 1 as it travels through curved space-time. This map is equivalent to a {displacement followed by a} beamsplitter as depicted in the lower diagram. It is nonlinear because $\xi, \gamma$ depend on the initial state, and because the initial state has to be ``copied'' to modes $3,4$, which violates the no-cloning theorem. In this diagram, there are two copies of the state $\rho^{(\textrm{in})}$; one acts on the modes $1,2$ and the other acts on the modes $3,4$. }
	\label{fig:EC}
\end{figure}

The event operator formalism is a non-linear extension to standard relativistic quantum field theory.
{Building upon the underlying theory presented in Refs.~\cite{RMD09,RAL14}, we present a detailed, case by case analysis of the model addressing the practical implementations to support a complete feasibility study.} To facilitate this we introduce an effective theory, applicable under the conditions of the proposed experiment.
Consider an initial two-mode Gaussian state  $\rho^{(\textrm{in})}_{1,2}$  (where $1,2$ label the modes) {generated by} the source on the ground. Mode $1$ is sent into space and detected on the ISS, while mode $2$ is detected on Earth  {at some small spacio-temporal displacement to ensure that the} two detection events are space-like separated. Under these conditions the event operator formalism can be represented by a map $\mathcal{E}$, between input $^{(in)}$ and final states $^{(fin)}$: $\rho^{(\textrm{in})}_{1,2} \mapsto \rho^{(\textrm{fin})}_{1,2}$. Unlike typical quantum channels, which are \emph{linear} completely positive trace-preserving (CPTP) maps, the event operator channel $\mathcal{E}$ is a fundamentally \emph{non-linear} CPTP map.
The map $\mathcal{E}$ {is equivalent to} {a displacement $D(\gamma)$ followed by} a beamsplitter with reflectivity $\xi$ (see Fig.~\ref{fig:EC}), where {$\gamma = \alpha_1 \,\frac{1-\sqrt{\xi}-\sqrt{1-\xi}}{\sqrt{1-\xi}}$} depends on the initial displacement $\alpha_1$ of mode 1, and where $\xi$ equals the ``event overlap''. The event overlap depends on the properties of the source mode, the intrinsic resolution of the detectors and the properties of the space-time along the optical paths. The full definition for $\xi$ is given in Ref.\cite{RAL14}. Below, we  give an expression for $\xi$ relevant for the proposed experiment.

We can use the equivalent optical circuit of Fig.~\ref{fig:EC} and the Schr{\"o}dinger picture to study the various cases and guide the design of the space mission.
In Fig.~\ref{fig:EC} the initial state is copied into the modes $3,4$: $\rho^{(\textrm{in})} \rightarrow \rho^{(\textrm{in})}_{1,2} \otimes \rho^{(\textrm{in})}_{3,4}$. 
The decoherence effect of gravity on mode $1$ is modelled by coupling it to its twin mode $3$ via the beamsplitter $\mathcal{E}$. If $A^{\dagger}_1,A^{\dagger}_3$ are photon creation operators for these modes, the beamsplitter evolution is given by a unitary $U_{\mathcal{E}}$ according to
\begin{eqnarray} \label{eventBS}
U_{\mathcal{E}} A^{\dagger}_1 U^{\dagger}_{\mathcal{E}} = \sqrt{\xi} A^{\dagger}_1 - \sqrt{1-\xi} A^{\dagger}_3 \nonumber \\
U_{\mathcal{E}} A^{\dagger}_3 U^{\dagger}_{\mathcal{E}} = \sqrt{\xi} A^{\dagger}_3 + \sqrt{1-\xi} A^{\dagger}_1 \,.
\end{eqnarray}
The state {experiences a displacement $D(\gamma)$ if $\gamma$ is nontrivial, and} evolves through the beamsplitter $\mathcal{E}$ using Eq.~\eqref{eventBS}. $\rho^{\textrm{(fin)}}_{1,2}$ is obtained by tracing out the modes $3,4$ and the expectation values for measurements are calculated. We now explore the consequences of this model for different input states.

\subsection{Two-mode time-energy entangled state from SPDC}\label{sec:timeent}

Consider an Spontaneous Parametric Down Conversion (SPDC) source with weak down conversion probability $|\chi|^2 \ll 1$. To a first-order approximation
the initial state is
\begin{eqnarray}
\ket{\psi}^{(\textrm{in})}_{1,2} \approx \ket{0} + \chi A^{\dagger}_1 A^{\dagger}_2 \ket{0} \,.
\label{pulse}
\end{eqnarray}
Copying this state to the modes $3,4$ we obtain
\begin{eqnarray}
&& \ket{\psi}^{(\textrm{in})}_{1,2} \otimes \ket{\psi}^{(\textrm{in})}_{3,4} = \ket{0} + \chi A^{\dagger}_1 A^{\dagger}_2 \ket{0} + \chi A^{\dagger}_3 A^{\dagger}_4 \ket{0} + \mathcal{O}(\chi^2) \,. \nonumber \\
&&
\end{eqnarray}
{Since the initial state is a squeezed vacuum: $\alpha_1 = \gamma=0$, the displacement does nothing.}
Using  the notation $A^{\dagger}_i A^{\dagger}_j \ket{0} := \ket{1, 1}_{i,j}$, we apply $U_{\mathcal{E}}$ to $\ket{\psi}^{(\textrm{in})}_{1,2} \otimes \ket{\psi}^{(\textrm{in})}_{3,4} $ to obtain:
\begin{eqnarray} \label{finalSqvac}
&& U_{\mathcal{E}} \ket{\psi}^{(\textrm{in})}_{1,2} \otimes \ket{\psi}^{(\textrm{in})}_{3,4}  \nonumber \\
&& = \ket{0} + \chi \sqrt{\xi} \ket{1, 1}_{1,2} -  \chi \sqrt{1-\xi} \ket{1, 1}_{3,2}  \nonumber \\
&& + \chi  \sqrt{\xi} \ket{1, 1}_{3,4} +  \chi \sqrt{1-\xi} \ket{1, 1}_{1,4} + \mathcal{O}(\chi^2). \,
\end{eqnarray}
Finally, by performing a partial trace over modes 3 and 4 we arrive at the final state 
\begin{eqnarray} \label{final}
\rho^{\textrm{(fin)}} &=&  (\ket{0} + \chi \sqrt{\xi} \ket{1, 1})( \langle 0 | + \chi \sqrt{\xi} \langle 1, 1 |) \nonumber\\
& +& \chi^2 (1-\xi)  (\densop{01}{01}+\densop{10}{10}) + \chi^2 \xi^2 \densop{0}{0}. \,\nonumber \\
\end{eqnarray}
To measure the coincidences, we apply the projector:
\begin{eqnarray} \label{detC}
\Pi_{C} = \iint \textrm{d}k \textrm{d}k' \,a^{\dagger}_{k,1} a^{\dagger}_{k',2} \densop{0}{0} a_{k,1} a_{k',2} \,,
\end{eqnarray}
which represents a coincidence measurement by ideal detectors that are frequency insensitive. By decomposing the mode operators into their spectral components we obtain $\langle \Pi_C \rangle = Tr\{\Pi_C \rho^{\textrm{(fin)}}\} \approx \xi \, |\chi|^2$, which is precisely the event operator prediction (compare to Eq.~(24) in Ref.~\cite{RAL14}). Also, note that there is no effect on the statistics of the individual modes (i.e.,\ singles),  as seen when we define:
\begin{eqnarray} \label{deti}
\Pi_{i} = \int \textrm{d}k \,a^{\dagger}_{k,i}  \densop{0}{0} a_{k,i} \,, \,\,i \in \{ 1,2 \} \,,
\end{eqnarray}
representing a single-photon detector for the $i^{\textrm{th}}$ mode, we find that $\langle \Pi_1 \rangle = \langle \Pi_2 \rangle  \approx |\chi|^2$. Hence the effect does not affect the singles counts.

The experimental implementation proposed here will only measure the decoherence of a time energy entangled state in the arrival time basis. Thus, the gravitationally induced decoherence can only be observed (with this experiment) as a decorrelation in the arrival times.
\subsubsection{Including losses}

A lossy channel can be modeled by a beamsplitter interacting with the vacuum state, whose transmission coefficient $0 <\eta<1$ equals the transmittivity of the channel~\cite{weedbrook2012gaussian}. Applying losses $\eta_1, \eta_2$ to modes $1,2$ results in the modified input state:
\begin{eqnarray}
&&\ket{\psi}^{(\textrm{in})}_{1,2} \otimes \ket{\psi}^{(\textrm{in})}_{3,4}\nonumber\\
&& = \ket{0} + \chi \sqrt{\eta_1 \eta_2} \,A^{\dagger}_1 A^{\dagger}_2 \ket{0} + \chi \sqrt{\eta_1 \eta_2} A^{\dagger}_3 A^{\dagger}_4 \ket{0} + \mathcal{O}(\chi^2) \,.\nonumber \\
\end{eqnarray}
After evolving through the $\xi$ beamsplitter, any further losses to mode $1$ is just another added factor that can be absorbed into $\eta_1$. The overall effect is just to transform $\chi \to \chi \sqrt{\eta_1 \eta_2}$.
This results in the coincidence rate:
\begin{eqnarray}
\langle \Pi_C \rangle \approx \eta_1 \eta_2 \xi \, |\chi|^2\,. \label{eq:loss}
\end{eqnarray}
In addition, since dark counts happen at the detectors, they won't change the factor $\xi$.

\subsubsection{CW operation}
Equation~\ref{pulse} describes a pulsed source producing spectrally uncorrelated photons, i.e.,\ the joint spectral amplitude for the source is separable, to first-order in $\chi$. However, in the experiment we propose to use a continuous wave (CW) source which inevitably implies strongly spectrally correlated photons. We can represent the initial state in this situation by
\begin{eqnarray}
\ket{\psi}^{(\textrm{in})}_{1,2} \approx \ket{0} + \chi \int \textrm{d}k H(k) a_{k,1}^{\dagger} a_{k,2}^{\dagger} \ket{0} \,,
\label{CW}
\end{eqnarray}
where, now the joint spectral amplitude -- H(k) is strongly correlated. Following the same procedure as before, i.e.,\ copying the state onto ancilla modes, interacting with the beamsplitter, tracing out the ancilla modes and modelling detection with a broadband detector, produces the same result as before, Eq.~\ref{eq:loss}.

\subsection{Coherent states}\label{sec:CoherentStates}

Suppose the initial state contains only classical correlations, in the form of two coherent states: $\ket{\psi}^{(\textrm{in})}= \ket{\alpha \beta}$. Now, the event operator map will apply a non-trivial displacement {$\gamma = \alpha \,\frac{1-\sqrt{\xi}-\sqrt{1-\xi}}{\sqrt{1-\xi}}$}. After copying the state to modes $3,4$ we obtain the following evolution:{
	\begin{eqnarray}
	\hspace*{-5mm}&&  U_{\mathcal{E}}D(\gamma) \,\ket{\alpha, \beta}_{1,2} \otimes \ket{\alpha, \beta}_{3,4} \nonumber \\
	\hspace*{-5mm}&& = U_{\mathcal{E}} \,\ket{\alpha, \beta}_{1,2} \otimes \ket{(\alpha+\gamma), \beta}_{3,4} \nonumber \\
	\hspace*{-5mm}&& = \ket{\sqrt{\xi}\alpha +\sqrt{1-\xi}(\alpha+\gamma),\beta}_{1,2} \nonumber \\
	\hspace*{-5mm}&& \otimes \ket{\sqrt{\xi}(\alpha+\gamma) -\sqrt{1-\xi}\alpha, \beta}_{3,4} \nonumber \\
	\hspace*{-5mm}&& = \ket{\alpha, \beta}_{1,2} \otimes \ket{\sqrt{\xi}(\alpha+\gamma) -\sqrt{1-\xi}\alpha, \beta}_{3,4} \,.
	\end{eqnarray}
}

After tracing out modes $3,4$ we are left with the same state we started with. This is just a special instance of the more general feature that classical correlations are preserved by the event operator formalism. Thus this theory predicts no decoherence with faint coherent pulses obtained, {for instance}, by attenuating a laser.

For non-Gaussian states (e.g.,\ classically correlated single-photons or photons from a deterministic single-photon source), the circuit of Fig.~\ref{fig:EC} fails to agree with the predictions of the event formalism, and it remains an open problem to find an accurate circuit that applies to these states. In this case, calculations performed directly in the event formalism confirm that classical correlations experience no gravitational {decoherence} (or {decorrelation}) effect in general.


\subsection{Polarization entangled SPDC states}

The de-correlation of entanglement due to event operators is not restricted to time-energy entanglement -- in principle it applies to any kind of entanglement. Let us consider the case of polarization entanglement with an initial state:
\begin{eqnarray}
\ket{\psi}^{(\textrm{in})}_{1,2} \approx \ket{0} + \frac{\chi}{\sqrt{2}} \ket{HH}_{1,2} + \frac{\chi}{\sqrt{2}} \ket{VV}_{1,2} \,,
\end{eqnarray}
using the notation e.g.,\ $\ket{HV}_{1,2} := A^{\dagger}_{H,1} A^{\dagger}_{V,2} \ket{0}$. After copying the state to modes $3,4$ and applying the beamsplitter, we obtain the state:
\begin{eqnarray}
\hspace*{-5mm}&& \ket{0} +  \sqrt{\xi} \,\frac{\chi}{\sqrt{2}} \left[ \ket{HH}_{1,2}+\ket{VV}_{1,2}+\ket{HH}_{3,4}+\ket{VV}_{3,4} \right] \nonumber \\
\hspace*{-5mm}&& + \sqrt{1-\xi} \,\frac{\chi}{\sqrt{2}} \left[ -\ket{HH}_{3,2}-\ket{VV}_{3,2}+\ket{HH}_{1,4}+\ket{VV}_{1,4} \right] \nonumber\\
\hspace*{-5mm}&& + \mathcal{O}(\chi^2) \,,
\end{eqnarray}
Computing the expectation values, we find that with probability $\xi |\chi|^2$ we obtain coincidences in which both photons have the same polarization. Coincidences in which the photons have different polarizations do occur, but only with probability $\approx |\chi|^4$, making these events negligible. So we cannot practically measure the de-correlation of polarization entanglement.

\subsection{Value of the event overlap}

Experimentally we are most interested in the case of the CW, time energy entangled SPDC. The event overlap for this case can be calculated using the methods described in Refs.~\cite{RMD09,RAL14}. If we assume that the spectral amplitude $H(k)$ is Gaussian, we obtain: 
\begin{eqnarray}
\xi = e^{- \kappa^2 / 2} \,,
\end{eqnarray}
where $\kappa^2 := \left( \frac{\Delta_t}{d_t} \right)^2$ is the dimensionless ratio of the gravitational time-dilation, $\Delta_t$, to the photon coherence time, $d_t$. The photon coherence time is the temporal standard deviation of $h(t)$, where $h(t)$ is the Fourier transform of the joint spectral amplitude --- $H(k)$, which is assumed to be a Gaussian. The gravitational time-dilation is the difference, $\Delta_t = t_d - \tau$, between the propagation times of the photons sent to the ISS as measured by local observers along the path, $\tau$, and a global observer situated far from the gravitating body $t_d$. We find
\begin{eqnarray}
\Delta_t = \int_{r_e}^{r_e+h} \textrm{d}r {{m}\over{r}}(1+{{2m}\over{r}}+{{r_e^2 \tan^2{\theta}}\over{r^2}})^{1/2} \,,
\end{eqnarray}
where $r_e$ is Earth radius, $h$ is the ISS height, $m$ is the mass of the Earth expressed in units of length\footnote{When working in natural units and in the context of gravitational physics or relativity it is common practice to express mass of a spherical body in terms of its Schwarzschild radius.} and $\theta$ is the angle from the zenith. This result is obtained assuming ${{m}\over{r}} \ll 1$. If we further assume ${{h}\over{r_e}} \ll 1$ and consider the result at the zenith ($\theta = 0$) we obtain $\Delta_t \approx {{m h}\over{r_e}}$, in agreement with Refs.~\cite{RMD09,RAL14}.

\subsection{Delay lines and space-like separation}

{
	At this juncture, we address an ambiguity in the formalism that has its roots in the long-standing measurement problem in quantum mechanics. What happens to the state from Eq.~\eqref{pulse} after a photon is detected in mode $2$? For all practical purposes, the state is said to have ``collapsed", resulting in a two-photon state (actually a one-photon state if the measurement is destructive, as it is here). However, different interpretations disagree about whether this apparent collapse is a physical process, or merely illusory. According to the many-worlds interpretation, for instance, there exists another branch of the wavefunction in which a photon was not detected in mode $2$, and hence the state is still vacuum in that branch. The total state is therefore expressed as an entangled state that includes the environment, containing the detector and the scientists observing the outcome:}

\begin{eqnarray} 
\ket{\psi}^{(\textrm{in})}_{1,2} \rightarrow \ket{0}\ket{E_0} + \chi A^{\dagger}_1 A^{\dagger}_2 \ket{0}\ket{E_1} +\mathcal{O}(\chi^2) \,,
\end{eqnarray}
where $\ket{E_0}$ indicates a quantum state of the environment in which no photon was detected and recorded by experimenters, while $\ket{E_1}$ indicates that a photon was detected in mode $2$ and recorded. This state is still a superposition of the vacuum and two-photon state for modes $1,2$, so it is still entangled in photon number. The event operator model therefore predicts a drop in coincidence counts due to the decorrelation of this entanglement.
{
	On the other hand, according to an objective collapse interpretation, the state after measurement of mode $2$ should be just the two-photon state:}

\begin{eqnarray} \label{instate}
\ket{\psi}^{(\textrm{in})}_{1,2} \rightarrow A^{\dagger}_1 A^{\dagger}_2 \ket{0} \,.
\end{eqnarray}
This state could still be entangled in its internal degrees of freedom (such as polarization) but it is clearly not entangled in photon number and hence the event operator model does not imply any loss of coincidences. We therefore face a dilemma: the predictions of the event operator model seem to depend on whether one uses a many-worlds or an objective collapse version of the model!

\begin{figure}[!ht]
	\includegraphics[width=0.85\columnwidth]{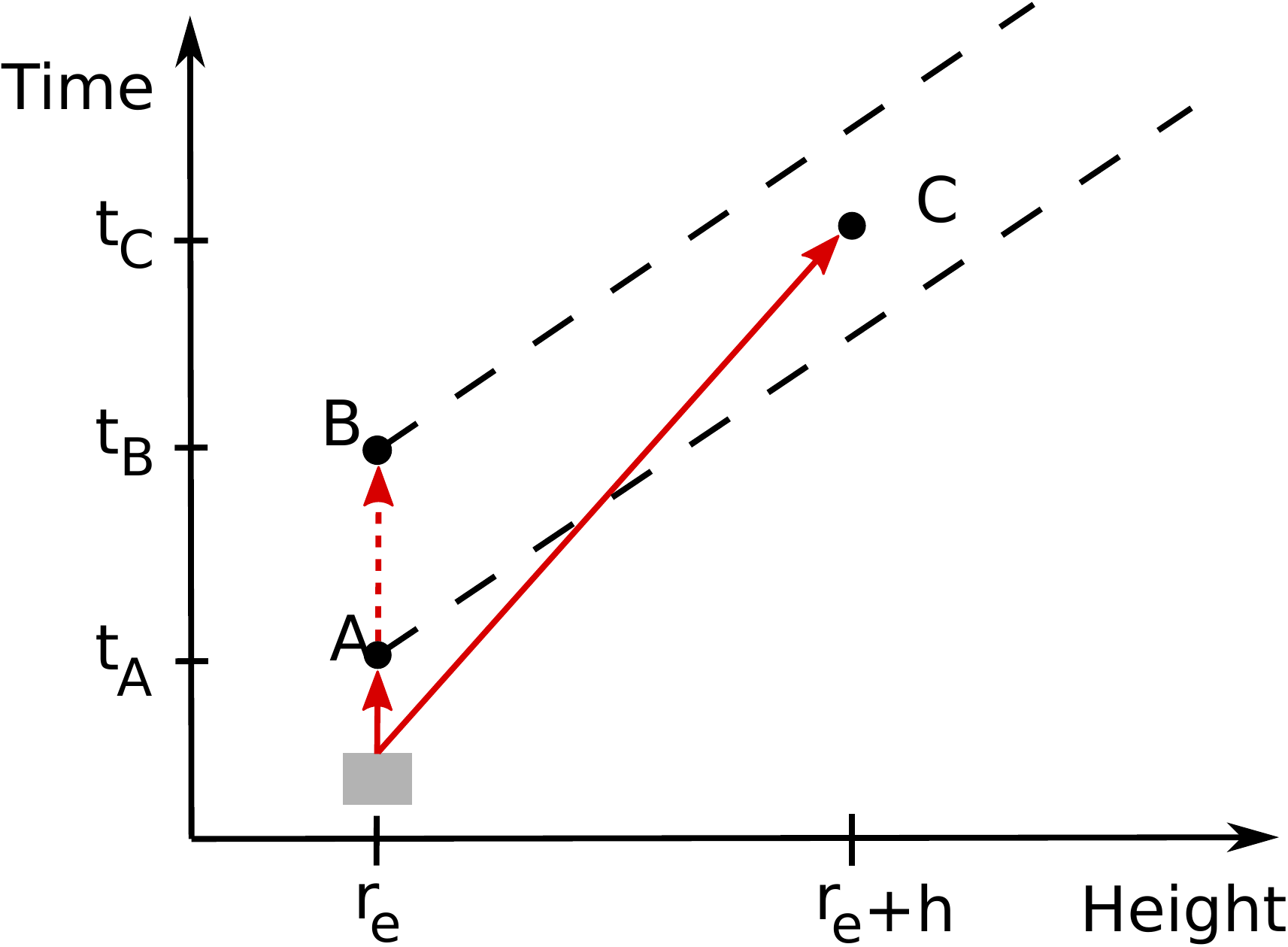}
	\caption{A space-time diagram showing the causal relationship between the
		detectors. The source (gray box) produces two photons, one of which is
		delayed on the ground and detected at time $t_A$ or $t_B$ (events $A$ and $B$)
		while the other (mode 1) is sent to space and detected at time $t_C$ (event $C$). The
		dotted lines indicate the path that would be taken by light traveling
		in a vacuum.
		As a result, the detection event $A$ is in the
		causal past of $C$, while event $B$ is causally separated from $C$. If the
		event operator model is modified to take into account the proposal of
		Kent~\cite{kent2005causal,KEN05}, then only photons detected at events $B$ and $C$ will experience gravitational decoherence.}
	\label{fig:cones}
\end{figure}

{
	Luckily, there is a very simple argument that shows that the objective-collapse version of the model must give the same predictions as the many-worlds version, in the special case of an experiment in which the heralding event (on Earth) is measured at space-like separation from the detection event (at the ISS), see Fig.~\ref{fig:cones}. The argument rests upon the empirical principle that no signals can be sent faster than light. Assuming that objective collapse interpretations must adhere to this principle, the objective collapse of the wave-function due to the detection of a photon in mode $2$, despite being instantaneous, cannot change the model's predictions from what they would have been if the two modes had still been entangled in photon number. If any such difference were permitted, it could be exploited to send a signal between space-like separated events, violating the no-signaling principle. (As an example of such a protocol, consider a state having photon number entanglement between three modes. The first mode is used to either collapse or not collapse the whole state, while the remaining two modes are used to check for the presence or absence of photon number correlations at a space-like separation). Non-signalling non-linear theories of this type were described by Kent in Refs.~\cite{kent2005causal,KEN05}.}
{
	The above argument shows that, in order to be consistent with the no-signaling principle, both versions of the event-operator model (the many-worlds version, and the `Kent' version) must predict a visible loss of coincidences when the detection events are space-like separated. Hence, space-like separation is necessary to conclusively test both variants of the model.
}
{
	We note that the Kent version of the model~\cite{KEN05} is also important to test because it has some advantages over the many-worlds variant. In particular, the many-worlds variant suffers from one aspect of the `preparation problem'~\cite{cavalcanti2012preparation} for non-linear theories, in that it does not make clear how to produce pure states operationally.  By constrast, the Kent version of the model  allows pure states to be created by measurement and post-selection, via an objective collapse of the wavefunction.}

Typically in QM a measurement is considered finished when the measurement has been stored as classical information that the quantum system can no longer affect. \footnote{We note that other interpretations require the motion of a sufficiently large mass~\cite{salart2008spacelike}}. This process takes a certain amount of time, typically on the order of 100 -- 200 ns. Thus on the ground the only delay needed is about 200\,m of optical fibre to obtain a delay of $\approx$1\,$\mu$s.

{Other than this small distance necessary to space-like separate the detection events, no other delay is needed. Let us consider an additional delay $\delta$ which is beyond that necessary for the aforementioned space-like separation. To see why $\delta$ does not play a role in the effect, we introduce $\delta$ to the case discussed in Section~\ref{sec:timeent}. This is done by applying the unitary $U_2(\delta)=e^{- i \delta a^{\dagger}_{k,2} a_{k,2}}$. Since it commutes with the beamsplitter and displacement of $\mathcal{E}$ (which don't act on mode 2), the state of all modes before detection is just:
	\begin{eqnarray}
	\hspace*{-5mm}&& U_{\mathcal{E}} \ket{\psi}^{(\textrm{in})}_{1,2} \otimes \ket{\psi}^{(\textrm{in})}_{3,4}  = \chi \sqrt{\xi} \,U_2(\delta) \,\ket{1, 1}_{1,2} + \,\,\textrm{(trivial terms)} \,. \nonumber \\
	\hspace*{-5mm}&&
	\end{eqnarray}
	The projector $\Pi_C$ is not sensitive to this phase shift, since:
	\begin{eqnarray}
	U^{\dagger}_2(\delta) \Pi_C U_2(\delta) &=& e^{ i \delta} \Pi_C e^{ - i \delta} \nonumber \\
	&=& \Pi_C \,,
	\end{eqnarray}
	hence the expectation value is unaffected.}

We emphasize that although the detection events need to be space-like separated, we do not need to make a fast basis choice as in a loophole free Bell experiment. We measure the decoherence effect by measuring in fixed measurement bases and looking at the change in correlations between measurement outcomes.

\section{Feasibility constraints}\label{sec:feas}
Section~\ref{sec:theory} as well as previous works~\cite{RAL14,Ralph2006,RMD09,Pienaar2016} predict an unusual and unexpected behavior of some systems. These predictions challenge standard quantum theory and pose interesting interpretations to relativity. Therefore it becomes very important to experimentally verify {these effects}. A successful detection of gravitational decoherence would have far reaching consequences for relativistic QM. On the contrary, 
proving the absence of (or experimentally imposing more stringent limits to) this decoherence effect can be used to test between several models in general relativity.

The mission needs to provide scientifically rigorous and  meaningful results and be practically possible. In this section we evaluate the feasibility of testing the above theory using a very simple single-photon detection module on board the ISS. In general, we show the feasibility in a worst case scenario, i.e.,\ we choose the worst alternatives/set of parameters that we can reasonably expect and for which the experiment remains possible.

\subsection{Quantifying the effect}

\begin{figure}[ht]
	\includegraphics[width=0.85\columnwidth]{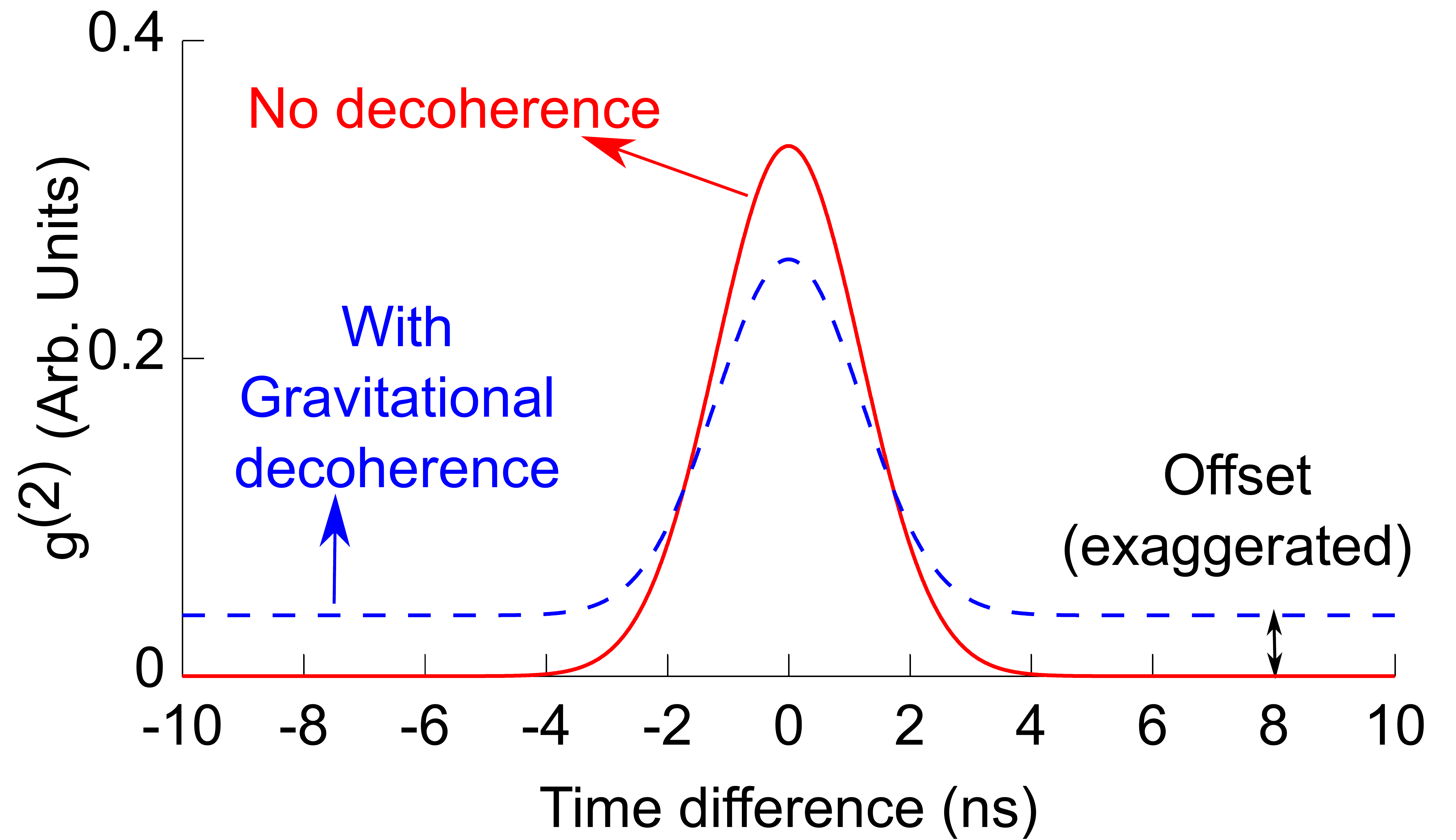} 
	\caption{Illustration of the gravitational decoherence effect. Consider a temporal cross correlation histogram $g^{(2)}$ between the arrival times of photons at the OGS and on the ISS. The area of the peak represents the number of photon pairs while the number of singles events is obtained from the photon counting module. The gravitational decoherence effect from Ref.~\cite{RAL14}, should result in a decrease in the number of photon pairs (area) without altering the singles rate, the position of or the width of the peak. This is depicted in the above figure where the red (solid) curve shows the $g^{(2)}$  in the absence of a gravitational field gradient (i.e.,\ without gravitational decoherence effect) and the blue (dashed) curve shows the effect of gravitational decoherence  between an OGS and the ISS at the zenith 400 km away using a source of time entangled photon pairs with a coherence time of 0.8\,ps. The offset shown here is grossly exaggerated and for illustrative purposes only. Therefore, to observe the gravitational decoherence effect we cannot rely on measuring the change in noise/background accidental count rates, instead we rely on measuring the change in area between the two curves. We emphasise that the gravitational decoherence effect can still be observed despite a detector jitter of several ns. Reducing the jitter only improves the Signal to Noise Ratio (SNR) by reducing the accidental coincidence rate (which contributes to the offset). 
		\label{fig:g2}
	}
\end{figure}

Section~\ref{sec:theory} shows that the effect is only present when using entangled states, further time-energy entanglement would produce a large observable effect unlike polarization entanglement. We have also demonstrated that to observe the effect it is sufficient to measure the decorrelation in one specific fixed  measurement basis. 
Consider a time-energy entangled state produced from a SPDC source (Section~\ref{sec:timeent}). The effect is observable as a reduction in the coincidence rate ($\langle \Pi_C \rangle$) without a reduction in the singles rates ($\langle \Pi_1 \rangle $ and $\langle \Pi_2 \rangle $). We define the heralding efficiency as $\frac{\langle \Pi_C \rangle}{\sqrt{\langle \Pi_1 \rangle \langle \Pi_2 \rangle}}$. From Eq.~\ref{eq:loss}, we observe that the change in heralding efficiency $E$ due to gravitational decoherence is given by the decoherence factor ($D_f$) defined as $D_f =  \eta_1 \eta_2 \xi $. Consequently, if $E_0$ is the heralding efficiency in the absence of any decoherence, the efficiency with the effect ($E_{\textrm{decoherence}}$) is given by: 
\begin{eqnarray}\label{eq:df}
E_{\textrm{decoherence}} = D_f  E_0\,.
\end{eqnarray}
This decoherence factor ($D_f $) is the same as $C_{total}$ in Ref.~\cite{RAL14} and
\begin{eqnarray}\label{eq:ctotal}
D_f \propto e^{-\frac{\Delta_t^2}{2d_t^2}}.
\end{eqnarray}
Experimentally, pairs (coincidences) are identified based on their individual arrival times~\cite{burnham1970observation} as recorded by independent time tagging modules. Typically, a cross correlation histogram ($g^{(2)}$) of these arrival times is used to identify coincidences. The width of the $g^{(2)}$ peak is limited by the detector jitter~\footnote{We note that the jitter in arrival time due to a highly turbulent atmosphere is $\ll$10\,ps and can be neglected compared to the $\approx$ 1--2\,ns jitter of the space based detectors $\approx$ 200\,ps combined jitter of the space based electronics, ground based electronics and ground based detectors.}, while accidental coincidences (noise) prevent the $g^{(2)}$ outside the peak from falling to zero. Figure~\ref{fig:g2} shows the expected change in the $g^{(2)}$ histogram with and without the gravitational decoherence effect. Photon pairs that undergo gravitational  decoherence lose correlation in their arrival times and (if not lost) are detected only as singles. Thus, in the $g^{(2)}$ histogram, they contribute only to the offset from zero. However, since the decohered pairs can contribute evenly to one of several time bins, the contribution to any one bin and thus to the offset is negligible in practice.

\subsection{Types of measurements}
We measure the decoherence of the time energy entangled photon pair by observing only in one degree of freedom --- the arrival time of each individual photon. Measuring in an energy degree of freedom would require a significantly more complex experiment. Thus, we can only observe what will appear to be a decorrelation effect. The largest challenge of this experiment is to distinguish the decorrelation from losses and background noise.
We plan to achieve this through a combination of \emph{four} different measurements out of which three depend on the functional dependence of $D_f$ with three different parameters and the last relies on comparison to a classical system that does not undergo decoherence.

Consider Eq.~\eqref{eq:ctotal}, $D_f$ depends on two factors we can change during  an experiment: The gravitational time dilation $\Delta_t$ depends on two parameters ---  the distance between the Optical Ground Station (OGS) and the ISS as well as the total gravitational potential difference. From the perspective of an observer on the ground these parameters can be expressed in terms of the zenith angle $\theta$ (i.e.,\ the angle~subtended by the ISS with the observer's zenith) and the orbital altitude $h$. Further, $D_f$ depends strongly on the coherence time ($d_t$) of the photons. Lastly, Section~\ref{sec:CoherentStates} predicts that $D_f = 0$ for a non-entangled faint pulse source (FPS); thus by comparing photons from a FPS and an entangled photon pair source (EPPS) we could detect the presence of gravitational decoherence.

\subsubsection{Variation of $D_f$ with the coherence time}
\begin{figure}[ht]
	\includegraphics[width=0.85\columnwidth]{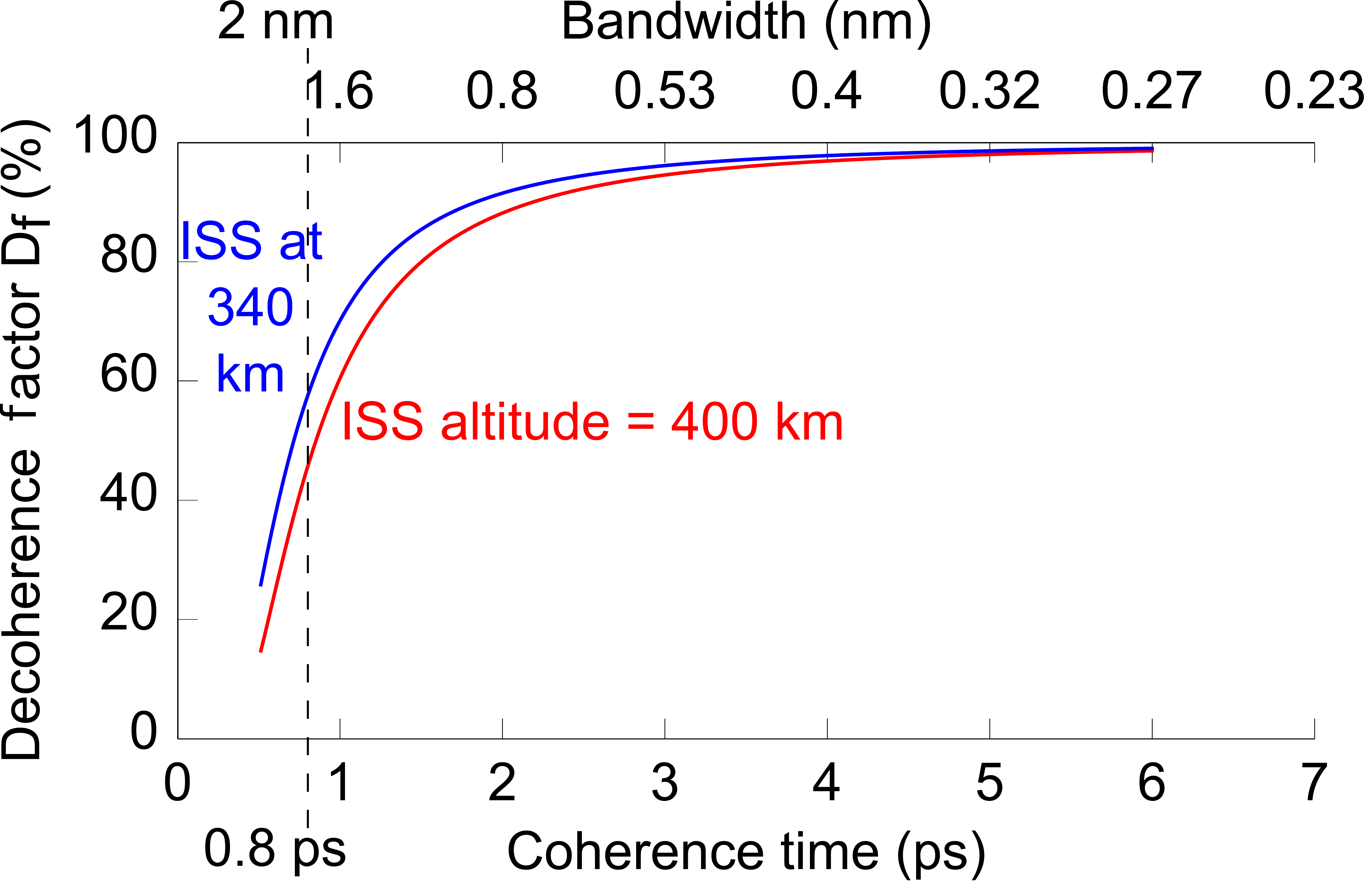} 
	\caption{Decoherence factor ($D_f$) as a function of the photon coherence time for two different orbital heights of the ISS.
		Ref.~\cite{RAL14} predicts that the dechoherence effect will be more pronounced for entangled photon pairs where the signal and idler modes have a large bandwidth (corresponding to a short coherence time). \label{fig:coherencetime}}
\end{figure}

The decoherence effect as quantified by $D_f$ is dependent on the coherence time of the photons used as shown in Fig.~\ref{fig:coherencetime}. 
Experimentally, varying the coherence time in the approximate range of 0.8\,ps to 3\,ps (i.e.,\ $\approx$2\,nm to 0.5\,nm bandwidth at 830\,nm) can be easily achieved by using a few different spectral filters. 
We recommend using this range as a good compromise between the increase in losses due to dispersion in the atmosphere, the strength of the effect, and the brightness of the source.

\subsubsection{Variation of $D_f$ with the orbital altitude}
\begin{figure}[ht]
	\includegraphics[width=0.85\columnwidth]{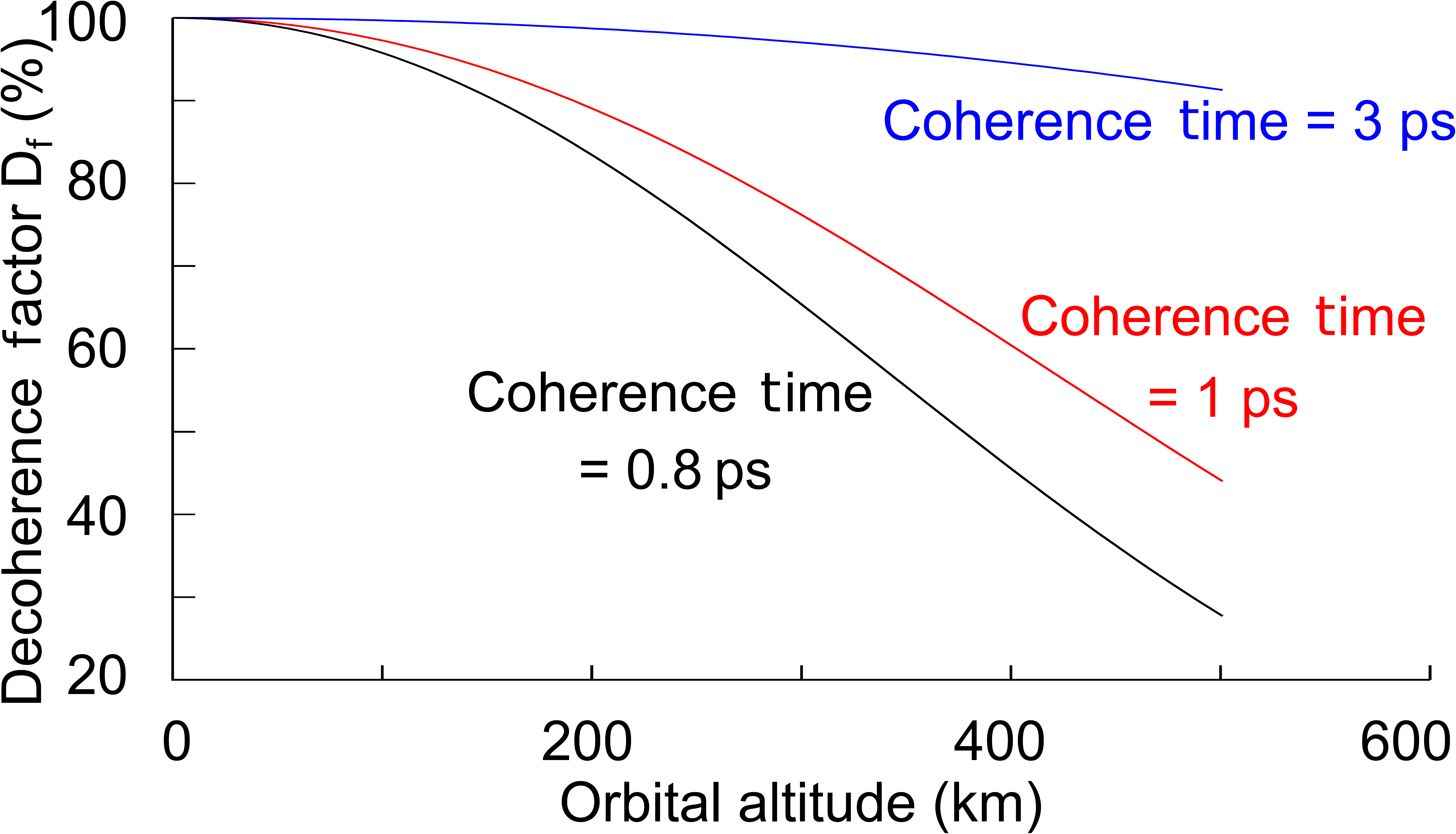} 
	\caption{
		Decoherence factor $D_f$ as a function of the orbital altitude $h$ of the ISS. These calculations were made according to Ref.~\cite{RAL14} for the case when the ISS is at the zenith of the OGS. The effect is more pronounced for broadband photons, which have a short coherence time.\label{fig:height}}
\end{figure}

By varying the orbital altitude of the ISS, we predict a large and measurable change in $D_f$ as seen in Fig.~\ref{fig:height}. Boosters on the ISS are used to control this orbital altitude. Currently the perigee altitude is maintained at 400 $\pm$ 2\,km. However this was not always so: in the years 1999 to 2009 the orbital altitude of the ISS underwent changes from $\approx$340 to 400 \,km at rates of more than 40\,km/year.

Operational constraints prevent any large changes in the orbital altitude.
In its current orbit, the ISS has an apogee of $\approx$412\,km and a perigee of 400\,km. This height difference is clearly insufficient to measure a significant change in $D_f$. Further, to exploit the ellipticity of the orbit  would require the ISS to have both its apogee and perigee at the zenith of the OGS. Even in this case, weather conditions and a limited mission lifetime (to further reduce costs) could hinder this measurement. Still, we believe this is important to consider the possibilities of performing such measurements on the ISS, or if unfeasible there, in future missions.

\subsubsection{Variation of $D_f$ with the zenith angle} 
\begin{figure}[ht]
	\includegraphics[width=0.85\columnwidth]{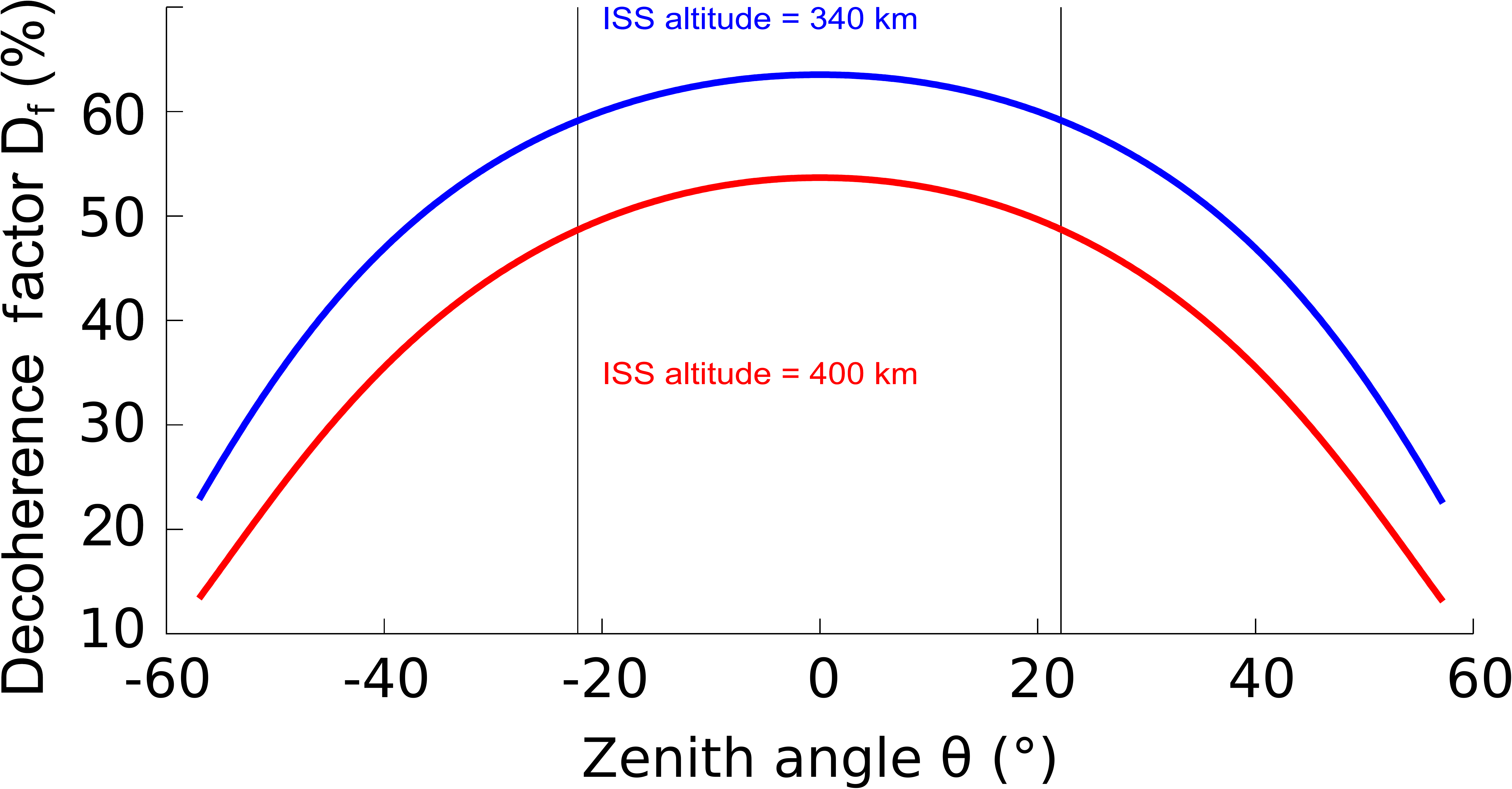} 
	\caption{The expected decoherence due to gravity is expected to be weakest at the zenith ($0^\circ$) and becomes stronger away from the zenith. Depending on the angle between the ISS and the OGS as well as the height of the ISS, we can predict the level of decoherence. The ``ISERV" telescope on board the ISS currently has a limited viewing angle of $\approx \pm22^\circ$ and this is represented here by the vertical black lines. \label{fig:angle} }
\end{figure}

Most passes of the ISS over a given OGS will be at some angle $\theta$ away from the zenith ($0^\circ$). This zenith angle also affects $D_f$ and must be taken into account (Fig.~\ref{fig:angle}). Telescopes on the ISS have a limited viewing angle represented by the vertical lines, Therefore even if the ISS is in the field of view of the OGS, the pass may not be usable. 

\subsubsection{Real time comparison of faint pulse versus entangled photons}
The previous three subsections have dealt with the characteristic ways in which $D_f$ changes when we change an experimentally controllable parameter. Another way to test the presence of gravitational decoherence of a quantum entangled system is to send an otherwise identical non-entangled photon along with one photon from an entangled pair and observe the difference between these two types of systems.   
A FPS with exactly the same wavelength as the EPPS can be used as an experimental control. The pulse width can be adjusted to produce attenuated single-photon states with exactly the same coherence times (for 830\,nm this corresponds to pulse widths of $\approx$340\,fs to 2\,ps).
By rapidly multiplexing photons from a FPS and an EPPS on a time scale much faster than the atmospheric turbulence, we can ensure a direct comparison. Naturally, to minimize error bars we must have high count rates, and the next section addresses this problem.

\subsection{Feasibility of the measurements}
To show the feasibility of the experiment we must focus on three aspects: the losses, the error bars due to counting statistics and the ability/sensitivity to perform the measurement despite the motion of the ISS.

\subsubsection{Losses}

Losses in a free space uplink to the ISS are {attributed to several different factors} --- absorption and scattering in the atmosphere, clipping losses due to a limited sending and receiving telescope apertures (the divergence of the beam due to diffraction limits and dispersion in the atmosphere further contributes to this loss),  beam wander and pointing accuracy that also cause clipping, limited detection efficiency, and limited transmission through the sending and receiving optics and telescopes.

\begin{figure}[ht]
	\includegraphics[width=0.85\columnwidth]{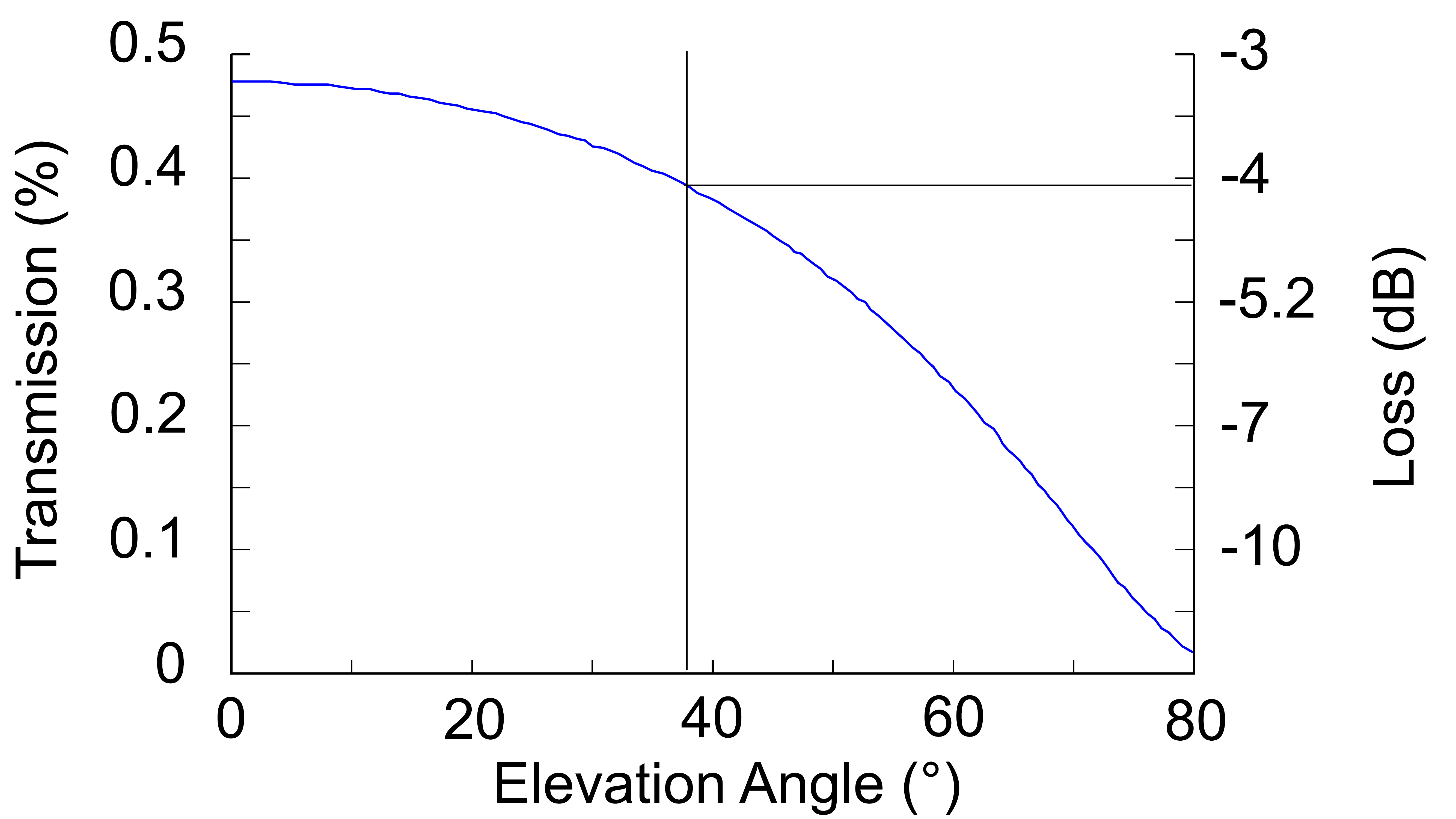} 
	\caption{Away from the zenith ($0^\circ$), a beam passes through a thicker column of air resulting in increased losses. The black line represents the maximum field of view of the NightPod from the Cupola window of the ISS. 
		Figure from Ref.~\cite{bourgoin2013comprehensive}.\label{fig:atmos}}
\end{figure}

First, let us consider atmospheric losses. These can be accurately modeled using the software MODTRAN 5 under a variety of weather conditions. We used the work of Ref.~\cite{bourgoin2013comprehensive} as a case study, choosing the OGS at Tenerife under  typical weather conditions of 20$^\circ$\,C, 50\% humidity, and a clear night. We chose a wavelength of 830 nm and model the losses from sea level (Fig.~\ref{fig:atmos}). In the worst case the transmission is better than $-$4.5\,dB ($\approx-$3.5\,dB near the zenith, i.e., in the best case).

Second, the clipping losses are due to the large size of the transmitted beam compared to the small receiving telescope. The size of the beam at the ISS depends on the distance to the ISS, the atmospheric turbulence  and the sending telescope used. Let us consider the worst case value of each of these: Given a maximum zenith angle $\theta$ of 37$^\circ$~\footnote{\label{note1}In Fig.~\ref{fig:angle} we used the limitation of the ISERV telescope because this gives the smallest change in $D_f$ (i.e.,\ the worst case scenario. Here we have chosen the field of view of the NightPod since that corresponds to the worst transmission.}
the distance to the ISS is $<$530\,km with a nominal orbital altitude of 400\,km. The Fried parameter $r_0$ (an indication of the size of pockets of turbulence in the atmosphere) can be used to determine the optimal sending telescope diameter for the smallest spot size at the ISS. For a typical OGS at Tenerife (say) we know that $r_0 \gg$ 15\,cm for most of the time~\cite{Frieddata}. Using the link budget application developed for ESA~\cite{linkbudgetapp}, we compute that the optimal sending telescope diameter is about 13\,cm. This will result in a diffraction limited spot diameter of $<$2.1\,m. Considering telescope imperfections, we expect a final beam diameter of $\approx$3\,m. To make a conservative worst case estimate let us consider a beam diameter of 4.5\,m. 

On board the ISS we have a choice of receiving telescopes --- a ``Nikon AF-S Nikkor 400\,mm" photographer's telephoto lens with a clear aperture of 13\,cm mounted in the Cupola window, and a 23.5\,cm astronomical telescope (Celestron CPC 925) mounted in the Window Observational Research Facility (WORF) window  (part of the ``ISERV" mission). On 12 May 2016 a tiny fragment of space debris caused a crack in the center of the Cupola window and operations were suspended pending repairs. For now we plan on using the 24\,cm telescope which gives us a clipping loss of between 26 to 28\,dB (calculated assuming a Gaussian beam profile)\footnote{These values are larger than the geometric estimate because the clear aperture could be slightly smaller than the specified diameter. Additional losses are due to the secondary mirror and its support structure.}.

Third, let us consider the beam wander loss which is limited by the pointing accuracy of the sending telescope (the receiving telescope can use fast mirrors to track the ground beacon). The pointing error of the OGS is measured against the position of a distant star and includes the atmospheric seeing (turbulence) effects and mechanical alignment. For example, the OGS at Tenerife has a minimum pointing error of 1.45\,$\mu$rad (with a fine adjustment mirror installed)~\cite{MeyerOGS}, thus it is reasonable to expect the pointing accuracy for other similar OGSs to be $<\approx$5\,$\mu$rad. Due to the fast motion of the ISS (up to $\approx$1.1$^\circ\per\second$), we must also consider about 5\,$\mu$rad of additional error in the point ahead angle. Thus, we have a total angular error of 10 to 15\,$\mu$rad.  To achieve this tracking precision even when the ISS is in the Earth's shadow, it will be necessary to equip both the OGS(s) and the ISS with tracking beacons. Using the results of Ref.~\cite{semenov2010entanglement} we estimate the beam wander loss to be $\approx$6\,dB.

Lastly, let us consider a 60\% detection efficiency in space, a 70\% transmission through the sending optics (and telescope) and the source, a 60\% (75\% in the best case) transmission through the multi-layered ISS window and a 70\% transmission through the receiving optics. All together the transmission for optics part of the uplink is $\approx-$7.5 dB

To estimate the total losses we combine the losses from each of the above to obtain the total worst case transmission as $-$46\,dB (Best case:  $\approx-$40\,dB).

\subsubsection{Fluctuation of count rates}

The experiment to detect gravitational decoherence using entanglement relies on the experimental capability of detecting  changes in the heralding efficiency.
In the absence of atmospheric turbulence, all non-systematic errors can be minimized by accumulating a large number of counts and averaging over several experimental runs. 
However, atmospheric turbulence influences both the signal count rate and the background count rate simultaneously thus averaging or accumulating statistics over long periods cannot reduce the error due to background count fluctuations. For a successful experiment we must identify the heralding efficiency change despite these fluctuations. Atmospheric turbulence occurs on the time scale of a few ms. We can rapidly alternate between sending photons from the FPS and from the EPPS on the time scale of  $\approx$100\,$\mu$s. The FPS photons would not undergo decoherence while the EPPS photons would. Thus to show that the decoherence effect occurs it would suffice to compare photons from these two sources. To show that this is feasible we shall consider the statistical distribution of fluctuations in the background and signal count rates as well as systematically varying losses that could exhibit the same behavior as the change in $D_f$ with the zenith angle $\theta$ (see Fig.~\ref{fig:angle}).

Typical photon counting statistics are Poissonian in nature. However due to atmospheric fluctuations in long distance links, the distribution is better modeled by the convolution of a Poissonian and a Log Normal Distribution (LND)~\cite{Capraro2012}. While the error of a Poissonian statistical sample scales as $\sqrt{N}$ for $N$ events, that of a LND scales as $s_i N$, where $s_i$ is the atmospheric scintillation index. Due to this unfavorable scaling it is very important to maximize the Signal to Noise Ratio (SNR) in the experiment.
In Low Earth Orbit (LEO), background counts can originate from several sources including detector's dark counts (see Appendix), direct sunlight, direct moonlight, light reflected from the atmosphere/clouds, light from on board the ISS and light reflected/emitted from the ground. In the worst case scenario we assume that all background counts originate on the ground and consequently follow a LND. Operating at night while the ISS is within the Earth's shadow and the moon is in a favorable position is essential to avoiding direct and reflected sunlight. Strong spectral filtering, shielding (optical and radiation), a small field of view and complete darkness surrounding the OGS can further reduce the background counts. 
By waiting for good weather conditions (i.e.,\ clear skies and $s_i < 0.05$ which is equivalent to a Fried parameter $r_0 > 28$\,cm) we can further reduce the effect of background counts. For high altitude observatories like those at Tenerife such suitable conditions occur 20 to 35\% of the time.

\begin{figure}[ht]
	\includegraphics[width=0.85\columnwidth]{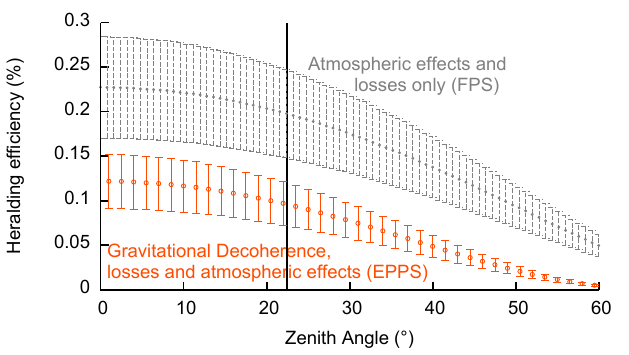} 
	\caption{Differentiating between the gravitational decoherence effect and losses due to the atmosphere: The gray curve (dashed) represent photons from the FPS and shows the result of simulations which only consider losses (due to the atmosphere, beam divergence, distance to the ISS, etc). The orange curve (solid) represents photons from the EPPS and shows the combined effect of the gray curve when we also consider the gravitational decoherence effect. The error bars are due to a log-normal error distribution of count rates caused by turbulence with a scintillation index of 0.05 with an accumulation time of 1\,s. The figure shows that we can still discern the presence of gravitationally induce decoherence despite a total of 6\,000 LND noise counts$\per\second$ on board the ISS. \label{fig:efferr}}
\end{figure}

Let us consider the change in the heralding efficiency due to the motion of the ISS from the zenith ($0^\circ$) towards the horizon ($90^\circ$) of the OGS. There are two main contributors to the change in the heralding efficiency: the change in losses (Fig.~\ref{fig:atmos}) and the change in the decoherence factor $D_f$ (Fig.~\ref{fig:angle}). Figure~\ref{fig:efferr} shows the expected dependence of the heralding efficiency on the zenith angle. The gray curve shows the predictions  which only takes into account losses and other atmospheric effects for the FPS (or of standard quantum theory).  The curve with orange error bars represents the combined effects of losses in the atmosphere and the gravitationally induced decoherence (i.e.,\ the change in $D_f$) predicted by Ref.~\cite{RAL14} for photons from the EPPS. The error bars represent 1 standard error in measuring the heralding efficiency using LND. The figure was computed using the worst case losses (46\,dB) and noise rates ($6000\per\second$ for both the space-based detectors all of which are assumed to have a LND). We note that the expected background rates are $< 1000\per\second$ in total i.e., $500\per\second$ for each of the two detectors and only the background counts can realistically be expected to have a LND. The remaining contribution to the noise count rate comes from the intrinsic dark counts which follow a Poissonian distribution. These dark count rates can realistically vary between $100\per\second$ to $2000\per\second$ depending on the amount of radiation damage to the detectors (see Appendix). Nevertheless, we conservatively assumed that all noise of the space based detectors follows a LND. We assume $200\,000$\,counts$\per\second$ (with a Poissonian distribution) divided among all detectors/pixels on the ground. We assume that both the FPS and EPPS each emit 350$\times$10$^6$\,photons$\per\second$ towards the space segment. The EPPS is assumed to have a 20\% intrinsic heralding efficiency. Thus on board the ISS we approximately expect 2650\,pairs$\per\second$ and 19\,500\,singles$\per\second$ inclusive of accidentals and noise counts most of which follow a LND.

The extent/strength of gravitational decoherence (if any) can be found by fitting the experimental data to either the gray or the orange curves in Fig.~\ref{fig:efferr}. To observe the decoherence effect it is sufficient to be able to differentiate between these two curves, which is still possible despite the significantly larger error bars of the LND. The curves shown here are for an orbital altitude of 400\,km. The atmospheric transmission losses are roughly the same for orbital altitudes between 300 to 500\,km. Only losses due to clipping change significantly, thus similar curves for different orbital altitudes will be parallel to each other.

Thus far in this subsection, we have considered distinguishing the gravitational decoherence  from losses and drastic/worst case fluctuations in background count rates at the level of one standard error. We note that the other methods of detecting gravitational decoherence discussed in this manuscript (such as, the strong variation with the coherence length of the photon pairs) are statically more rigorous and can lead to identifying the presence or absence of the gravitational decoherence by 6 or more standard errors. Further, measurements under more favourable weather conditions ($s_i < 0.05$) or with lower noise count rates would also increase the statistical significance of the results. Thus, scientifically meaningful conclusions can be drawn from the mission despite the limited statistical significance of one type of measurement. Nonetheless, designing the mission to ensure a statistical significance of 6 standard errors for the measurement described in this subsection would prohibitively increase the cost of the mission. 

\subsubsection{Sensitivity of measurements}\label{sec:sen}

The dependence of $D_f$ on coherence time, orbital altitude and zenith angle $\theta$ can be calculated theoretically and is shown in Figs.~\ref{fig:coherencetime},~\ref{fig:height} and~\ref{fig:angle}. In each of these cases we need to measure a change in $D_f$ ($\delta_{D_f}$) between say two positions of the ISS, two different coherence times or types of sources. The minimum value of $\delta_{D_f}$ we can resolve experimentally with one standard error is a function of  the signal and noise count rates as well as $s_i$. For a given $s_i = 0.05$ and a worst case loss estimate of 46\,dB, we numerically vary the production rate of entangled photon pairs and compute the maximum noise count rate that will enable us to still resolve a certain value of $\delta_{D_f}$. 

\begin{figure}[ht]
	\includegraphics[width=0.85\columnwidth]{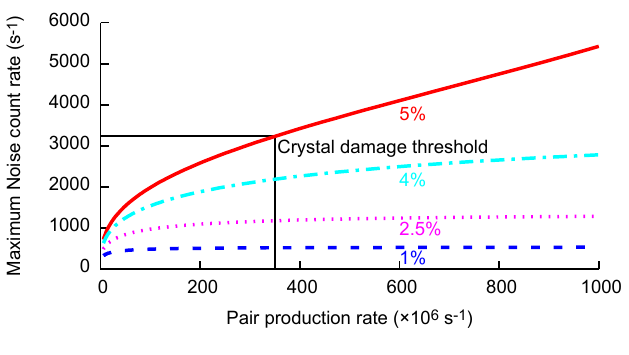} 
	\caption{{The maximum noise count rate we can tolerate for each of the two detectors on board the ISS using $s_i$ = 0.05 and the worst case transmission estimate of -46\,dB as a function of  rate at which entangled photon pairs are produced on the ground. The four different curves shown are the upper bounds for being able to resolve changes  in $D_f$ (i.e.,\ $\delta_{D_f}$) of 0.05, 0.04, 0.025 and 0.01 with at least one standard deviation significance.} \label{fig:pvsnoise}}
\end{figure}

Out of the 4 measurements we can perform to verify the decoherence effect, $D_f$ is least sensitive to changes in the zenith angle $\theta$ specially near 0$^\circ$. Nevertheless, we can show that even this measurement is feasible despite the large and worst case background count rates and fluctuations.
So far there have been no measurements of the background count rates we can expect using single-photon detectors in space with a narrow field of view. This makes it very difficult to estimate the background count rates we will observe in the final experiment. Our best estimates predict between 1000 to 5000 counts per second.  
The maximum field of view (MFOV) of the telescope through the WORF window on the ISS depends on the details of how it is mounted and how much room there is for the telescope to move. In the worst case the MFOV is limited to 45$^\circ$. Thus the maximum observable change in $D_f$ is from a zenith angle $\theta$ of  0$^\circ$ to  22.5$^\circ$. Here $\delta_{D_f}$ is 0.051 or approximately 5\% (see Fig.~\ref{fig:angle}). As seen in Fig.~\ref{fig:pvsnoise} we can tolerate up to 6000 noise counts per second and still be able to resolve this change. We emphasize that this is only a worst case estimate and the actual experiment can be expected to be much more sensitive because only a small fraction of the background light will follow a LND due to atmospheric turbulence as we expect the largest contribution to be light reflected from clouds, the upper atmosphere, or the ISS itself.

Similarly a 5\% $\delta_{D_f}$ can be obtained by changing the coherence time from 0.8\,ps to 0.864\,ps which corresponds to a decrease in the bandwidth by 0.14\,nm. It can also be obtained by varying the orbital altitude of the ISS by $\approx$31\,km
Thus in the worst case (i.e.,\ with a noise count rate of 2000$\per\second$ for each of the two ISS based detectors), we will be able to detect the effect if we were to change the bandwidth by about 0.16\,nm, the altitude by $\approx$31\,km, or the zenith angle $\theta$ by  22.5$^\circ$.

The sensitivity of our measurements to a change in $D_f$ is strongly dependent on the noise count rate(as seen in Fig.~\ref{fig:pvsnoise}). The noise count rate consists of the background counts and dark counts. The former is roughly constant throughout the mission duration while the latter increases over time due to radiation damage to the detectors. The appendix details this effect and shows the maximum tolerable background count rate at various mission durations.
Decreasing the noise count rate to 950$\per\second$ on each detector allows us to be sensitive to a change in $D_f$ of 2.5\%. Thus the smallest change in orbital height that could be used to detect the decoherence effect is about 15\,km. Obtained when the orbital altitude changes from 400 to 415\,km at the zenith of the OGS when using a coherence time of 0.8\,ps (see Fig~\ref{fig:height}). Similarly, the smallest change in bandwidth that results in the smallest measurable change in $D_f$ of 2.5\% is about 0.08\,nm, obtained by changing the coherence time from 0.8 to 0.84\,ps at an orbital altitude of 400\,km and at the zenith of the OGS.

We can clearly see that the best possible measurements of $D_f$ are to study the variation with coherence time  (due to its sensitivity) and to make a comparison of the EPPS with a FPS.

\section{Proposed Experiment}\label{sec:expt}

\begin{figure*}[ht]
	\includegraphics[width=1.75\columnwidth]{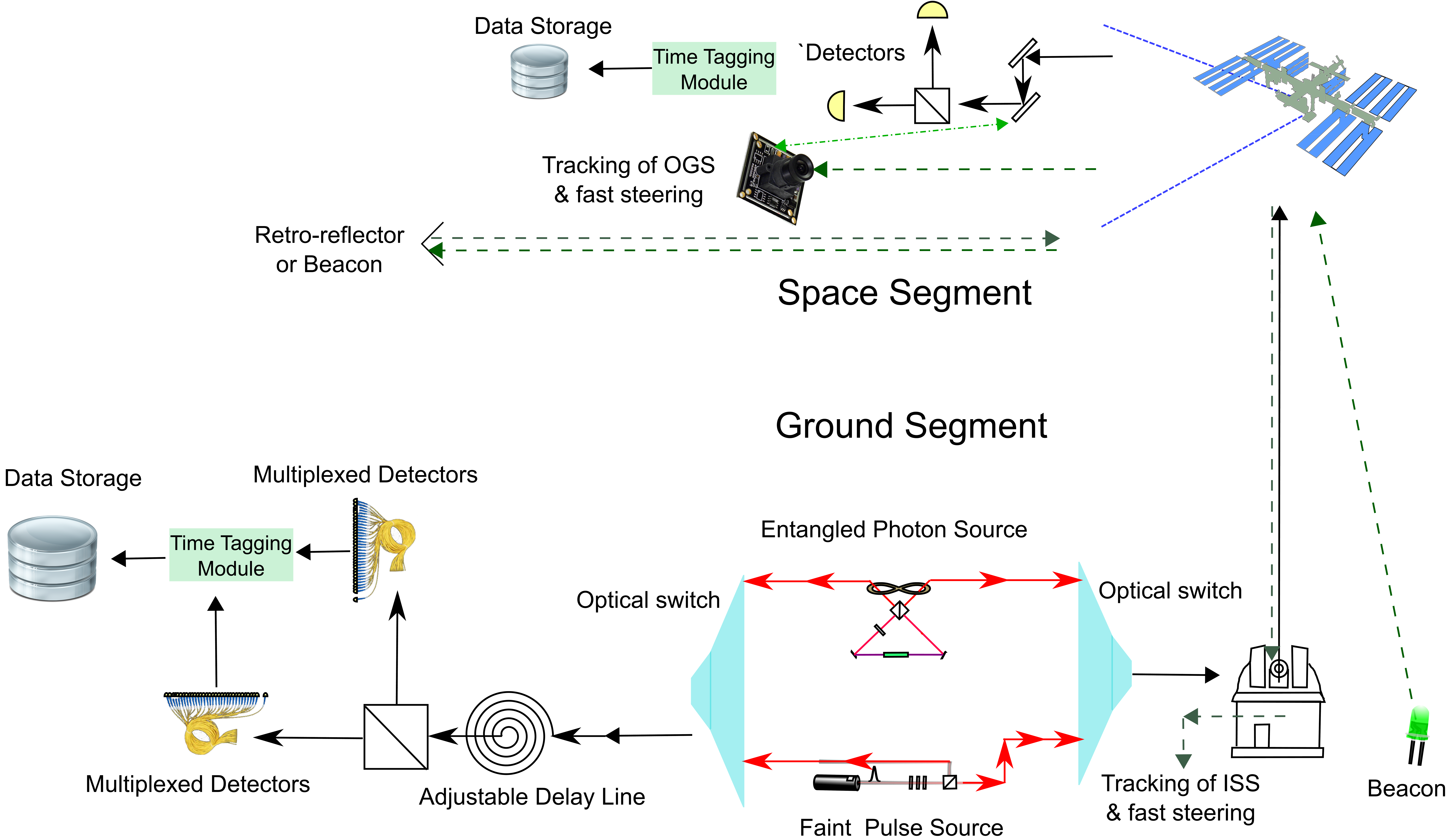} 
	\caption{{Schematic overview of the proposed experiment.}
		The OGS has a source of time and polarization entangled photon pairs. One photon of each pair is measured immediately after the source (i.e.,\ in the same gravitational potential), while the other photon propagates through a large gravitational field gradient before being detected on board the ISS. The arrival time of each photon is recorded and used to identify photon pairs in post processing. One photon of each entangled pair traverses a significantly different gravitational space-time metric. According to Refs.~\cite{RAL14,RMD09}, this should result in a decoherence-like phenomenon, which is measurable as a reduction in the number of pairs without a reduction in the number of singles.\label{fig:setup}
	}
\end{figure*}

%
The previous sections have shown how to calculate the effect and that it is experimentally measurable. In this section we discuss details about the experimental realization of Space QUEST. 
Fig.~\ref{fig:setup} shows a schematic overview of the experiment as well as key features of the OGS. Weather conditions and the flight path of the ISS prevent a single OGS from making the most of this experiment. Thus, it is advantageous to have several OGSs, for which there are several suitable candidates including but not limited to the Tide/Iza\~{n}a observatories at Tenerife~\cite{Meyer,MeyerOGS}, Matera Laser Ranging Observatory (MLRO)~\cite{OGSMatera}, Optical Ground Station Oberpfaffenhofen (OGSOP)~\cite{OGSOP} and  transportable optical ground station (TOGS)~\cite{TOGS,moll2015ground}, many of which have already been used for quantum experiments~\cite{schmitt2007experimental,OGSOP,villoresi2008experimental,ursin2009space}.
To enable quantum communication and Bell test experiments the ISS module includes two detectors and a Polarizing Beam Splitter (PBS) while the EPPS on ground is also capable of producing polarization entangled states.

The ISS (with a limited field of view of $\pm 37^\circ$ in the best case) and an OGS can maintain an optical link for about 10 to 300\,s during a usable pass. During this time, the OGS and ISS should acquire each other and commence tracking,  perform housekeeping/maintenance measurements and then start the experiment. We can divide each experiment into several integration time windows of say 1\,s each, during which calibration measurements are followed by the quantum experiment. We suggest the following utilization pattern for each integration time window:
\begin{itemize}
	\item 5\% for measuring the intrinsic dark counts of the detectors by using a shutter to block all incident light.
	\item 15\% for measuring the background count rate by blocking the transmission of optical signals at the OGS.
	\item 10\% for measuring the optical link loss by sending pulses of known intensity. The measurement of time delays, clock synchronization and polarization distortions can be performed in this time window by controlling the duration, timing, and polarization of the calibration pulses.
	\item 29\% of the time for experiments with the FPS (`classical system').
	\item 40\% of the time for experiments with the EPPS (`quantum entangled system').
	\item 1\%  for switching between the various modes.
\end{itemize}

\begin{figure}[ht]
	\includegraphics[width=0.85\columnwidth]{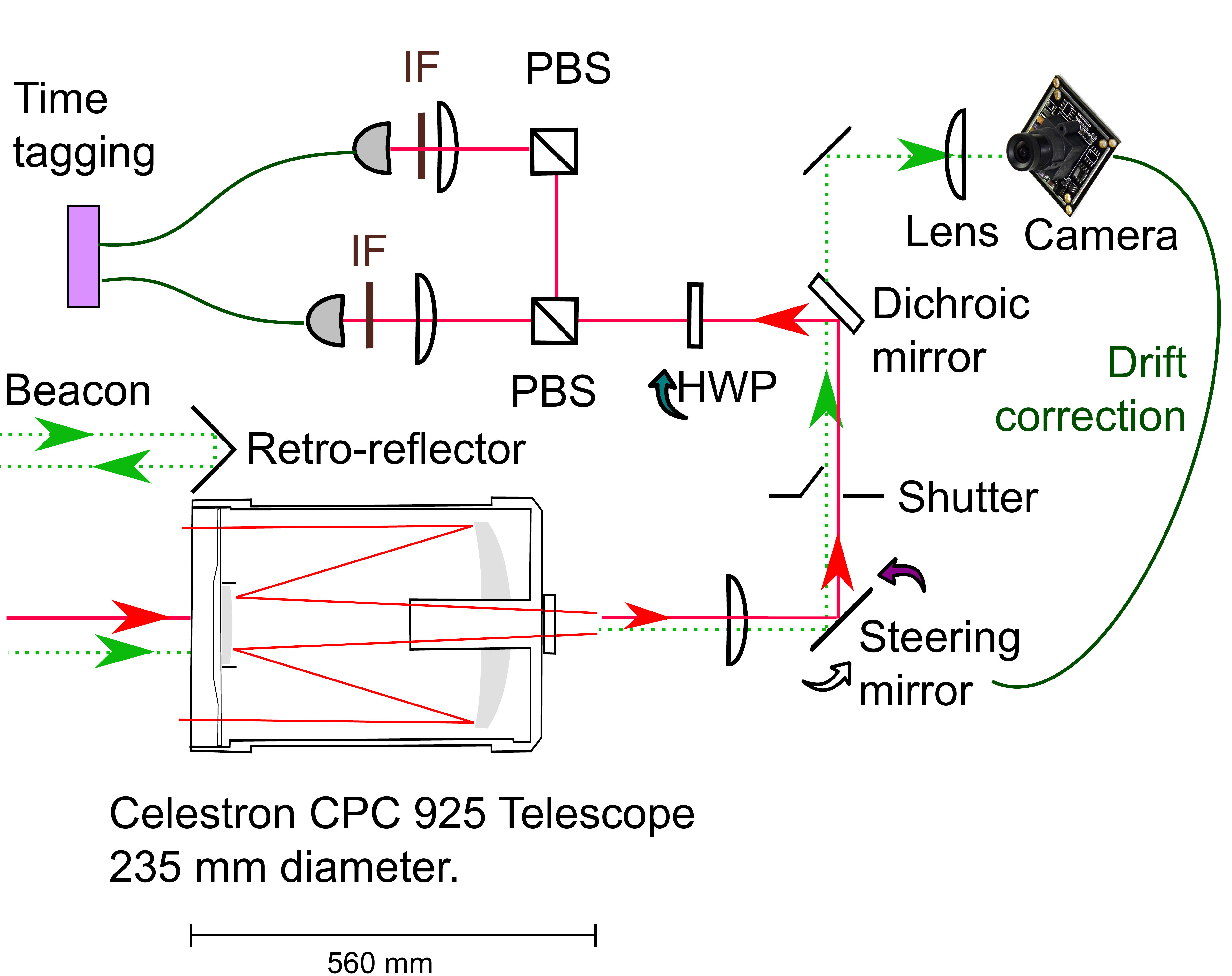} 
	\caption{{Schematic of the proposed ISS segment of the experiment.}
		This scheme shows the key elements of the experimental setup that will need to be on board the ISS. We plan to use a standard 23.5\,cm hobby astronomy telescope system as the receiving telescope. The single-photon signal is isolated from the beacon by a dichroic mirror and sent to a polarization analysis module. A time tagging module records the arrival times of each single-photon signal as well as the polarization information. A lens system images the beacon onto a camera. A computer uses the camera image to control a steering mirror to actively compensate for any pointing inaccuracies of the telescope mount and the ISS. A retro-reflector can be used to reflect the OGS's tracking beacon and enable fine tracking of the ISS in the Earth's shadow.\label{fig:issseg} The Half Wave Plate (HWP) is used to change the polarization measurement basis and the Interference Filter (IF) helps eliminate unwanted background counts by limiting the spectral sensitivity of the detectors.}
\end{figure}

One possible implementation of the experimental setup on board the ISS is shown in Fig.~\ref{fig:issseg}. The single-photon signals are collected by a receiving telescope (mounted facing the Earth and capable of tracking the OGS), separated by a polarization analysis module (consisting of an adjustable Half Wave Plate (HWP) and a PBS), and detected by single-photon detectors (with a jitter $<$ 2\,ns) after they pass through narrowband Interference Filters (IF), which remove the majority of background noise. Time tagging electronics record the arrival time of each photon with a resolution of $\approx$100\,ps. A beacon laser emitted by the OGS is used for tracking. The laser is separated from the single-photon signal by a dichroic mirror and  detected using a camera. An optional steering mirror can be used for fine tracking. An optical shutter is necessary to prevent damage to the detectors due to bright light. Lastly, a retro-reflector for the beacon (or a second beacon laser), mounted near the receiving telescope, enables the OGS to track the receiver. 

The bulk of the setup in space consists of the receiving telescope. Fortunately we can use the existing telescope (``ISERV") from the ``SERVIR"  mission on board the ISS --- which consists of a stable automatic tracking mechanism for photographing the Earth's surface, as well as a 23.5\,cm diameter Schmidt-Cassegrain telescope both of which are currently installed in the Earth facing WORF window of the ISS. The remainder of the minimalistic setup shown in Fig.~\ref{fig:issseg} will be built as a compact attachment to the eye-piece of the receiver. Si avalanche photodiodes (APDs) with thermoelectric cooling can be used for photon detection (see Appendix). The time tagged information can be used in conjunction with its counterpart on the ground to identify pairs and look for gravitational decoherence. Meanwhile, the polarization information can be used to verify the quantum nature of the system via Bell tests and perform quantum communication.

On the other hand,the OGS shall be capable of:
\begin{itemize}
	\item Quickly (in $\approx$0.1\,ms) switching between different sources or blocking the output.
	\item Providing an adjustable delay (of up to $\approx$ 0.1 to 1\,$\mu$s) to ensure a space-like separation of detection events.
	\item Measuring and storing the data from the extremely high pair production rates. This could be achieved using arrays of detectors. We estimate that $\approx$ 390\,TB of data will need to be stored (over a half year mission duration). 
	\item Compensating for polarization drifts after dedicated calibration measurements. The multi-layered, thick windows on the ISS may cause angle dependent polarization rotations.

\end{itemize}
All of these requirements can be accomplished with existing technology.
The biggest challenge is the production of a few hundred million photon pairs per second. We currently have a type 0 Periodically Poled Potassium Titanyl Phosphate (PPKTP) based source capable of producing 8$\times10^6$\,photon pairs$\per\second$/mW of pump power. Increasing the pump power would in principle allow us to generate at the required pair rates. The ground based detectors should be capable of detecting a singles rate of $>1.5\times10^9$\,photons$\per\second$. This can be done by multiplexing several detectors. In principle single-photon detector arrays like Si APD arrays with $\gtrapprox$200\,pixels or nanowire arrays with $\approx$16 to 32\,pixels could be built which would be ideal candidates. For example, the lunar laser communication demonstration used 4 arrays of 4 nanowire detectors~\cite{nanoarray} and APD arrays are commercially available~\cite{apdarrayidq}. Such systems can be adapted/combined for use in the OGS.
The ground and spaced based detection schemes together should be able to correctly identify photon pairs. Which means that the bin width (limited by the timing jitter of the electronics) must be much smaller than the mean time between local detection events. Thus a jitter of better than 225\,ps would be sufficient to ensure that the probability that two photons arrive in the same bin is less than 0.05 (assuming Poissonian statistics).Further, we estimate that the total system detection efficiency of the multiplexed detectors should be better than 20\%.


Similarly the very simple ISS segment shown in Fig.~\ref{fig:issseg} shall be able to:
\begin{itemize}
	\item Measure the arrival time of photons with a resolution of 100\,ps~\footnote{
		The time tagging resolution, i.e.,\ digitization bin width, should be much better than the detector jitter (specified earlier as $<$ 2\,ns) to avoid additionally broadening coincidence peak and to facilitate accurate clock synchronization by means of accurately measuring the timing position of the coincidence peak}
	as well as measure the incoming photons in a selectable linear polarization basis. The detectors should be capable of measuring up to 250\,000\,photons$\per\second$~\footnote{The corresponding pair rate as seen between the ISS and ground is estimated to be $\approx$50\,000\,paris$\per\second$} (The maximum expected rate when the OGS produces 300$\times 10^6$\,pairs$\per\second$).
	\item Time synchronization of the Space QUEST clock on board the ISS with the OGS clock to better than 100\,ns~\footnote{Using a clock, on board the ISS, synchronised to within a few tens of ps of the OGS clock is prohibitively expensive. Instead we can exploit the strong time correlations of photon pairs or that of the FPS to correlate the time tags between the ISS and OGS in post processing.}.
	\item Store the $\approx$2\,TB of data (generated over a mission duration of half a year). We note that all data analysis is done in post processing.
	On board measurements are not needed in real time and can be provided on a hard drive at the end of the mission. The near real-time data transfer can be limited to housekeeping and calibration/verification data thus reducing the load on the limited communication bandwidth of the ISS.
	\item Track and maintain the OGS within the field of view of the detectors. For the schematic shown in Fig.~\ref{fig:issseg} a 500\,$\mu$m diameter of the active area would be sufficient for tracking given an atmospheric scintillation index $<$ 0.1. 
\end{itemize}

The requirements of the ISS segment can be  met with existing commercial technology. 

\section{Discussion and Conclusions} \label{sec:end}
In this paper we have evaluated various methods to measure the gravitational decoherence effect. We have identified the best and most scientifically rigorous way forward while using existing commercially available technologies, studied the feasibility of the scheme and identified key requirements and hurdles towards implementing this experiment in a ground to ISS uplink scenario.
We have shown the comparitive simplicity of the end-to-end system, with most of the complexity on the ground, as well as the feasibility of the experiment for the ISS.

QM, being a \emph{linear} theory, predicts the absence of any gravitational decoherence {in the proposed experiment}. Consequently, if an effect was observed this would be a monumental achievement {that would overturn the traditional view about how quantum matter interacts with the gravitational field}. Nevertheless,
should our experiment fail to detect gravitational decoherence, a first upper bound will be established in a benchmark experiment.   This limit will provide direct experimental evidence to bound the possible non-linearity of QM {in the presence of gravity}, and allow us to place bounds on the maximum decoherence predicted by various models {of gravitationally-induced decoherence}.
{The absence of decoherence in this experiment would suggest that QM should not be modified in order to conform to the predictions of classical general relativity. One motivation for the event-operator model is consistency with non-hyperbolic space-times such as closed time-like curves. Thus a plausible conclusion in this case is that general relativity (GR) has to be modified to accommodate (linear) QM, in one of the following possible ways:}
\begin{enumerate}
	\item {It may be that the non-linearity only manifests in cases where the local curvature is due to the presence of a Closed Time-like Curve (CTC) somewhere in space-time (and not due to, e.g.,\ a massive planet). However,} this would imply that local physics can depend on the global topology of space-time, in violation of the equivalence principle.
	\item {Physical laws (such as unknown quantum gravity effects) might prevent CTCs from existing at all, which would remove the motivation for the non-linear model considered here. However, since CTCs are a direct prediction of GR, this option would clearly require a modification of GR}.
	\item It may be that CTCs can exist, and their nonlinear effects can be observed in general curved space-times, but that the nonlinearity is described by a model other than the event-operator model, for example a field theory extension of Post-selected Closed Time-like Curves (P-CTCs) {along the lines of the path integral approaches discussed in Refs.~\cite{POL81,BOUL,FRIE}}. In this case, the experiment would place bounds on the size of this non-linearity.
	\item Finally, it could be that the event-operator model is correct, but decoherence is not observed because all correlations are fundamentally classical (i.e.,\ entanglement is really the result of a classical realistic hidden-variable theory). Due to Bell\textsc{\char13}s theorem, this would imply reality is non-local, which is arguably contrary to the local structure of GR.
\end{enumerate}
The only one of the above options that does not immediately require a modification of GR is option (3). However, there is evidence that P-CTCs and possibly other CTC models would imply the ability to signal information {between events that are not causally connected} in the space-time metric, also violating a basic principle of GR. Although it might be possible to find a model that does not have this pathological feature, nobody has yet seen how to achieve this despite much effort, making it unlikely to be the case.

Thus, the significance of the experiment {can be summarized as deciding whether QM becomes non-linear in the presence of gravity (in which case decoherence is predicted), or whether the theory of GR will ultimately need to be changed in order to allow for the linearity of QM (if entanglement is seen to be preserved).}

To have a scientific payload outside the ISS is more demanding than locating it inside the ISS~\cite{ursin2009space}. The approach proposed in this paper significantly reduces the financial burden but is constrained by safety requirements applicable to internal payloads. Still, it can be implemented, making use of already available infrastructures and hardware. In addition, it has a very low-cost and it can be easily upgraded.

The primary objective of the Space QUEST mission is to search for the gravitational decoherence effect. However, the secondary objectives of the mission include quantum communication in an uplink between the ground and the ISS. The setup we proposed here would be more than sufficient to achieve this as it exceeds the requirements given in Ref.~\cite{Scheidl2013a}.

In this paper, we have shown that the experiment is feasible and discussed its scientific importance for all possible outcomes. Furthermore, all technologies, instruments and other requirements of the mission are readily achievable using existing commercially available products.  Several key components needed by this experiment are already on board the ISS~\cite{nightpod,ISERV}, which drastically reduces the cost of the proposed mission. We strongly believe that this experiment or similar needs to be undertaken to resolve the above mentioned scientific conundrums.

\section{acknowledgments}
	This work was funded by ESA under the grant number:
	20772/07/NL/VJ (Topical Team).
	We  would like to thank all members of the ESA topical team for their suggestions. 
	We specially thank our counterparts at ESA  Olivier Minster, Nadine Boersma and Zoran Sodnik. Finally we wish to thank Norbert L\"{u}tkenhaus and Cesare Barbieri for their support and contributions to the Space QUEST mission.  VP acknowledges financial support from the Spanish Ministry of Economy and Competitiveness through the “Severo Ochoa” program for Centres of Excellence in R\&D (SEV-2015-0522), from Fundació Privada Cellex, and from Generalitat de Catalunya through the CERCA program


\appendix*
\section{Feasibility of using Si APDs as single-photon detectors in the ISS segment}
\label{sec:feasibility-detector}

Silicon avalanche photodiodes (APDs) are good candidates for detectors in the ISS segment, because they have good photon detection characteristics around $830~\nano\meter$, are well understood, widely used in quantum optics, and do not require deep cryogenic cooling \cite{cova2004}. However, proton radiation present in low Earth orbit damages the APDs and drastically increases their dark count rate over time~\cite{sun1997,sun2001,sun2004,tan2013,tang2016}. Here we simulate the radiation environment inside the ISS to study the feasibility of using Si APDs. Although the mission duration is expected to be less than a year, we calculate for a 2-year exposure to have a contingency margin. 

\begin{figure}[h!]
	\centering
	\includegraphics{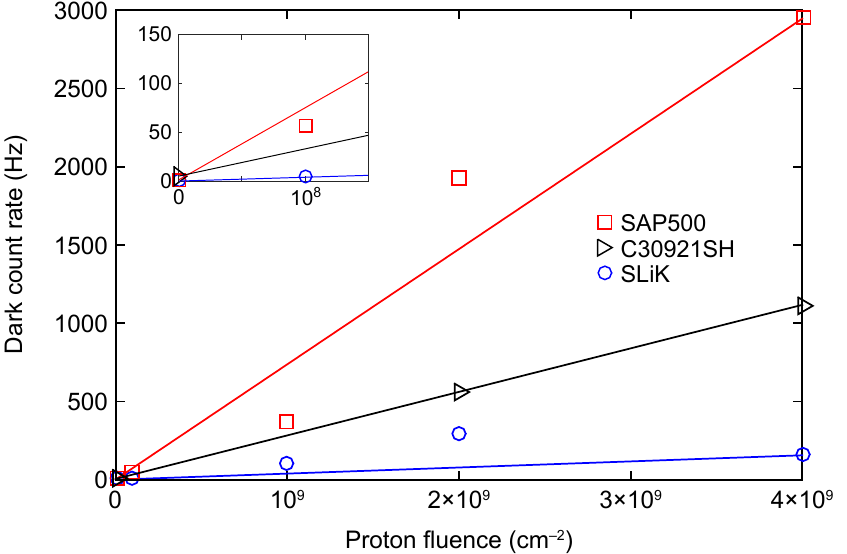}
	\caption{\label{fig:dark-count-rate-irradiation}{\bf Dark count rates of three different APD models at various radiation fluences at $\bm{-86~\celsius}$} \cite{anisimova2017}. Each data point is an average between the two samples tested. Linear fit lines pass through the first and last data points. The dark count rate increase appears to be linearly proportional to the proton fluence. Inset is a close-up of the first points.}
\end{figure}

We use SPENVIS radiation simulation software for the ISS orbit (51.64$^\circ$ inclination, $401~\kilo\meter$ perigee, $409~\kilo\meter$ apogee, and 15.54 orbits per day). For proton radiation flux, we assume the solar minimum, because the solar cycle 24 will be at the solar minimum in 2018--2020, which gives the worst radiation damage to the APDs~\footnote{Increased UV radiation and solar activity at solar maximum cause the Earth's atmosphere to expand which removes trapped protons in the radiation belts ~\cite{solar-cycle-endnote-citation}.}. We assume that the detector module includes $20~\milli\meter$ thick spherical aluminum shielding. Storing it in a random place inside the ISS typically adds $10~\milli\meter$ further shielding by the pressure vessel and micro-meteoroid orbital debris impact shield of the ISS~\cite{christiansen2003,christiansen2009}. We thus simulate for a total of $30~\milli\meter$ spherical aluminum shielding. Displacement damage dose (DDD) after 2 years under these assumptions is $1.27\times10^{6}~\mega\electronvolt\per\gram$. While the simulated DDD monotonically decreases with increased shield thickness, it does not depend on it strongly for thicknesses that can reasonably be used in this mission. E.g.,\ doubling the total thickness to $60~\milli\meter$ would less than halve the DDD, while adding significant extra weight to the detector module.

We base our dark count rate estimates on proton irradiation tests reported in Ref.~\cite{anisimova2017}. Three different commercial models of thick-junction Si APDs were tested: Excelitas SLiK, Excelitas C30921SH, and Laser Components SAP500. The samples were irradiated by monochromatic $100~\mega\electronvolt$ proton beam, at fluences of $10^{8}$, $10^{9}$, $2\times10^{9}$, and $4\times10^{9}~\centi\meter^{-2}$. Two samples of each model were irradiated at each fluence, then their dark count rates measured at $20~\volt$ overvoltage \footnote{For Si APDs, detector performance weakly depends on the overvoltage. Detection efficiency and dark count rate increase, while jitter decreases at higher overvoltage~\cite{cova2004}.} and several temperatures down to $-86~\celsius$ (see Fig.~\ref{fig:dark-count-rate-irradiation}). The increase of the dark count rate appeared to be roughly linear on the fluence, although some sample-to-sample variation was observed, up to a factor of 3. Unpublished data at $-60~\celsius$ yielded similar conclusions.

\begin{table}[t!]
	\centering
	\caption{\label{tab:temperature-estimation}{\bf Estimated APD temperature required to reach various dark count rates after 2 years in orbit.}}
	\renewcommand{\arraystretch}{1.2}
	\begin{tabular}{
			>{\centering\arraybackslash}m{0.23\linewidth}
			>{\centering\arraybackslash}m{0.23\linewidth}
			>{\centering\arraybackslash}m{0.23\linewidth}
			>{\centering\arraybackslash}m{0.23\linewidth}}
		\hline\hline
		\textbf{APD model} & \textbf{$\bm{200~\hertz}$} & \textbf{$\bm{660~\hertz}$} & \textbf{$\bm{2000~\hertz}$} \\
		\hline
		SLiK & $-57.4~\celsius$ & $-42.7~\celsius$ & $-29.1~\celsius$ \\
		C30921SH & $-81.5~\celsius$ & $-65.1~\celsius$ & $-49.8~\celsius$ \\
		SAP500 & $-95.6~\celsius$ & $-77.1~\celsius$ & $-59.9~\celsius$ \\
		\hline\hline
	\end{tabular}
\end{table}

Our DDD value calculated above is equivalent to $5\times10^{8}~\centi\meter^{-2}$ at $100~\mega\electronvolt$ monochromatic proton fluence in the above test. Taking into account exponential dependence of the dark count rate on temperature \cite{anisimova2017}, we estimate the APD temperature required to reach the dark count rates of $200$, $660$, and $2000~\hertz$ at the end of the 2-year mission. The results are listed in Table~\ref{tab:temperature-estimation}. However, sample-to-sample variations and uncertainty of radiation environment prediction \cite{tang2016} necessitate a reserve factor. We assume the detector design needs to be able to cope with factor of 3 worse dark count rates than predicted, which requires cooling by an additional $\approx 15~\celsius$. Thus, to guarantee dark count rate below $2000~\hertz$ per APD, the detector module should be capable of cooling SLiK APDs to $-45~\celsius$ and C30921SH to $-65~\celsius$. At these temperatures, afterpulsing probability of SLiK and C30921SH is projected to stay below $1\%$ \cite{anisimova2017}. These temperatures are achievable with thermoelectric cooling and forced-convection air radiator at room temperature \cite{kim2011}. However cooling below $-65~\celsius$ may require a more complex design, possibly using a compact Stirling cooler. If a sufficient cooling system cannot be provided, implementing additional radiation damage mitigation methods can be considered, such as in-orbit thermal annealing~\cite{anisimova2017} or laser annealing~\cite{lim2017}.

\begin{figure}[t!]
	\centering
	\includegraphics[width = \columnwidth]{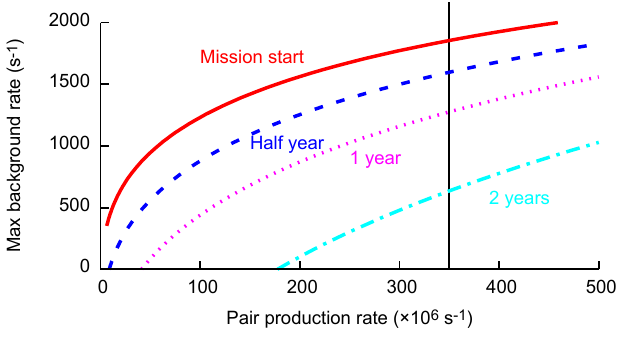}
	\caption{\label{fig:dark_maxbkgrnd}Maximum tolerable background count rates per SLiK detector, cooled to $-29.1~\celsius$, in the presence of radiation damage. The curves show the maximum background count rates per detector that would still allow us to resolve changes  in $D_f$ (i.e.,\ $\delta_{D_f}$) of 4\% with at least one standard deviation significance. At the start of the mission undamaged detectors have $<$ 100\,Hz dark counts while after two years they are expected to have dark counts of $\approx$ 2000\,Hz.}
\end{figure}

Both the dark count rate and the background count rate play a vital role in the mission feasibility. Due to the increase in dark count rates from radiation damage the maximum acceptable background rate decreases over time. 
Figure~\ref{fig:dark_maxbkgrnd} shows the maximum tolerable background count rate for the SLiK detector cooled to $-29.1~\celsius$, for various mission durations. This is calculated in the same way as Fig.~\ref{fig:pvsnoise} using the expected dark count rates at the end of various mission durations. As seen from Table~\ref{tab:temperature-estimation} cooling the detector further can drastically reduce the dark counts. This would also improve the sensitivity at which this mission can measure the gravitational decoherence effect. We predict a background count rate of about 500$\per\second$ for each detector. After 2 years of radiation exposure the measurements of gravitational decoherence would only be possible for pair production rates  $> 300\times10^6\per\second$ and would rapidly become impossible for longer mission durations. We note that it is unlikely that the Space QUEST mission duration would exceed 6 months to a year. Nevertheless to maintain significant safety margins,we recommend that the gravitational decoherence experiments be conducted in the early states of the mission and the secondary objectives be attempted later on. Further, most secondary objectives like quantum communication, light pollution measurements, etc. can still be successful with much larger dark count rates.

In summary, commercial thick-junction Si APD chips from Excelitas (SLiK) appear to be a suitable choice for the ISS segment, especially given that our science experiments can tolerate dark count rate of $1000$--$2000~\hertz$ per detector. The detector module will need to use a custom thermal design and electronics \cite{kim2011,pugh2017}. The noise budget presented in this paper already accounts for noise rates of up to 3000$\per\second$ per space based detector, of which we conservatively estimate that 500count$\per\second$ can be attributed to the background count rate. Thus the mission is feasible with the minimal radiation shielding provided by the ISS module. Further shielding could help increase the sensitivity of the decoherence measurement.

\bibliographystyle{plain}     

\end{document}